\documentclass[twocolumn]{aastex701}
\usepackage[utf8]{inputenc}
\usepackage{placeins}
\usepackage{float}
\usepackage{amsmath}
\usepackage{ulem}
\usepackage{hyperref}
\usepackage{CJK}
\defcitealias{Cashman24}{C24}
\defcitealias{Lelli14}{L14}
\defcitealias{Karachentsev04}{K04}
\defcitealias{Murphy14}{MB14}



\begin{document}
\begin{CJK*}{UTF8}{gbsn}
\title{Redshifted Cool Gas in the Disturbed H\,{\sc i} Disk of the Starburst Dwarf NGC\,4214}
\correspondingauthor{Taotao Fang}
\email{fangt@xmu.edu.cn}
\author[orcid=0000-0002-3940-2950]{Shulan Yan (鄢淑澜)}
\email{yansl@stu.xmu.edu.cn}
\affiliation{Department of Astronomy, Xiamen University, Xiamen, Fujian 361005, People's Republic of China}
\author[orcid=0000-0002-2853-3808]{Taotao Fang (方陶陶)}
\email{fangt@xmu.edu.cn}
\affiliation{Department of Astronomy, Xiamen University, Xiamen, Fujian 361005, People's Republic of China}
\author[orcid=0000-0002-3878-5590]{Xiaochuan Jiang (姜小川)}
\email{jiangxc@fjbu.edu.cn}
\affiliation{School of Information Engineering, Fujian Business University, Fuzhou, Fujian 350506, People's Republic of China}
\author[orcid=0000-0003-4874-0369]{Junfeng Wang (王俊峰)}
\email{jfwang@xmu.edu.cn}
\affiliation{Department of Astronomy, Xiamen University, Xiamen, Fujian 361005, People's Republic of China}
\author[orcid=0000-0003-3230-3981]{Qingzheng Yu (余清正)}
\email{qingzheng.yu@unifi.it}
\affiliation{Dipartimento di Fisica e Astronomia, Universit\`a degli Studi di Firenze, Via G. Sansone 1, 50019 Sesto Fiorentino, Firenze, Italy}
\affiliation{INAF - Osservatorio Astrofisico di Arcetri, Largo E. Fermi 5, I-50125 Firenze, Italy}
\author[orcid=0000-0002-4428-3183]{Chuanpeng Zhang (张传朋)}
\email{cpzhang@nao.cas.cn}
\affiliation{National Astronomical Observatories, Chinese Academy of Sciences (NAOC), Beijing 100101, People's Republic of China}
\affiliation{Guizhou Radio Astronomical Observatory, Guiyang 550025, People's Republic of China}
\author[orcid=0000-0001-6083-956X]{Ming Zhu (朱明)}
\email{mz@nao.cas.cn}
\affiliation{National Astronomical Observatories, Chinese Academy of Sciences (NAOC), Beijing 100101, People's Republic of China}
\affiliation{Guizhou Radio Astronomical Observatory, Guiyang 550025, People's Republic of China}
\affiliation{Guizhou Provincial Key Laboratory of Medium and Low Frequency Radio Astronomy Technology and Application, Guiyang 550025, People's Republic of China}
\begin{abstract}
We report redshifted cool gas toward the central star-forming region of the nearby starburst dwarf galaxy NGC\,4214. HST/COS spectra show low-ionization absorbers redshifted relative to the systemic velocity, most clearly in S\,{\sc ii} ($\Delta v=19.8\pm2.6\ {\rm km\,s^{-1}}$) and supported by Mg\,{\sc ii} ($\Delta v=30.3\pm3.0\ {\rm km\,s^{-1}}$), while P\,{\sc ii} remains tentative ($\Delta v=7.5\pm3.4\ {\rm km\,s^{-1}}$). The Keck/HIRES spectra also reveal a redshifted Ca\,{\sc ii} absorption component ($\Delta v=25.6\pm1.5\ {\rm km\,s^{-1}}$). Because the absorption is detected against the central UV or optical continuum, the gas must lie in the foreground of the star-forming region and is moving away from the observer and toward the background UV source in projection. A VLA--THINGS H\,{\sc i} spectrum extracted at the COS position within one VLA beam also shows a redshifted neutral component at $v_{\rm hel}=329.2\pm0.9\ {\rm km\ s^{-1}}$, offset by $38.2\pm1.3\ {\rm km\ s^{-1}}$ from systemic and by $27.6\pm1.0\ {\rm km\ s^{-1}}$ from the dominant local H\,{\sc i} component. The FAST H\,{\sc i} mapping further reveals an extended, kinematically disturbed neutral disk, with residual structures relative to a warped-disk model. Together, the redshifted UV absorption, the redshifted H\,{\sc i} component, and the disturbed large-scale H\,{\sc i} velocity field show that the central anomalous cool gas is observed in a galaxy with a disturbed extended neutral disk. This geometry suggests that the redshifted cool gas may trace inflow or recycled fallback toward the central region, although warped-disk or non-circular disk motions cannot be fully excluded.
\end{abstract}

\section{Introduction}
Dwarf galaxies are crucial for understanding the formation and evolution of the first galaxies in the universe \citep{Putman21}. The origin of intense star formation in these low-mass galaxies remains unclear. Gas migration in the past $\sim1\ \mathrm{Gyr}$ may enhance the formation of stars \citep{Hunter04,Tang22}. Torques due to massive star-forming clumps can make outskirt gas lose angular momentum and fall into the central region \citep{Elmegreen12}. In addition to the inner process, interactions \citep{Bekki08,Kimbro21,Chhatkuli23} and gas accretion from the intergalactic medium (IGM) \citep{Silk87,Egorov21,Lin25} can also feed and enhance star formation. \\

For dwarf galaxies with recent starbursts, the H\,{\sc i} disk tends to be disturbed \citep{Cannon04,Kobulnicky08,Lelli14b,Lelli14}. The extended and asymmetric H\,{\sc i} structures are mainly from the recent interactions or gas accretion \citep{Stil02,Lelli12,Lopez12,Chandola23}. The shallow potential well of dwarf galaxies makes it easier to expel gas out of the stellar disk. Most metals of these galaxies may reside in the circumgalactic medium (CGM) or the IGM \citep{Shen14,McQuinn15,Christensen18,Qu22,Zheng24}. On the other hand, the re-accreted gas from the CGM might be metal-rich \citep{Giavalisco11,Ribaudo11,Rubin12}. Therefore, starburst dwarf galaxies provide useful laboratories for studying gas cycling in low-mass systems. \\

NGC\,4214 is a nearby irregular dwarf starburst galaxy, which has a stellar mass of $1.5\times10^9\ M_\odot$ \citep{Karachentsev04} and star formation rate (SFR) of $0.13\ M_\odot\ \rm yr^{-1}$ \citep{Lelli14b}. From NASA/IPAC Extragalactic Database (NED), the distance is $3.0\pm0.3\ \rm Mpc$ and the heliocentric velocity is $291\pm1\ \rm km\ s^{-1}$. \citet{Sanchez23} revealed that it has young superbubbles with dynamical age $\sim2\ \mathrm{Myr}$ and supernova remnants with age $\sim10^4\ \mathrm{yr}$, indicating that NGC\,4214 has recent star formation. \citet{Walter08} observed NGC\,4214 in the H\,{\sc i} Nearby Galaxy Survey (THINGS) using the Very Large Array (VLA). Previous observations have shown that NGC\,4214 has an H\,{\sc i} disk extending beyond the stellar disk \citep{Walter08,Warren12}. Therefore, NGC\,4214 is a useful laboratory for investigating the connection between gas accretion and star formation. However, the limited sensitivity of the VLA data makes it difficult to trace more diffuse and extended H\,{\sc i} structures. \\

The Five-hundred-meter Aperture Spherical radio Telescope (FAST) is a powerful single-dish telescope in the northern sky. The FAST all sky H\,{\sc i} survey (FASHI) project has a median detection sensitivity of $\sim0.76\ \mathrm{mJy\ beam^{-1}}$ at a velocity resolution of $6.4\ \mathrm{km\ s^{-1}}$ \citep{Jiang19,Jiang20,Zhang24}. It can detect H\,{\sc i} sources with column density down to $10^{17}\ \mathrm{cm^{-2}}$ \citep{Xu25,Zhang26}, which can trace more extended H\,{\sc i} structures in galaxies. Apart from the global gas traced by H\,{\sc i}, the low-ionized gas traced by the ultraviolet (UV) absorbers is essential to understand the metals and feedback process \citep{Davies20,Qu22}. The Cosmic Origins Spectrograph (COS) is a UV spectrograph of the Hubble Space Telescope (HST). It provides a wavelength coverage from $1150$-$3200$\AA \citep{James14}. The grating G$130$M of COS covers the $900$-$1450$\AA with resolution ($\lambda/\Delta \lambda$) higher than $\sim10000$. It has an advantage in detecting low ionized gas (e.g., Mg\,{\sc ii}, S\,{\sc ii}), which can trace the cool gas ($T\sim10^4\ \mathrm{K}$) of the galaxies \citep{Gnat07,Lehner13,Werk14,Tumlinson17}. \\

Multiwavelength observations are essential for studying the baryon cycle in NGC\,4214. However, the connection between the central star-forming region and the extended gas reservoir remains unclear. Previous down-the-barrel studies of NGC\,4214 using FUSE, Keck/HIRES, and HST/COS reported broad or outflowing absorption but did not identify a comparable narrow redshifted component \citep{Schwartz04,Grimes09,Chisholm15}. We also examined the archival FUSE spectra. The extended-source FUSE LWRS spectrum lacks the effective resolution required to isolate a component offset by only $20$--$30\ {\rm km\,s^{-1}}$. \\


In this work, we combine HST/COS metal absorption lines with VLA--THINGS and FASHI H\,{\sc i} data. We use the VLA data to examine the neutral gas along the COS sightline, and the FAST data to trace the extended H\,{\sc i} disk. Our goal is to test whether the redshifted cool gas toward the central region is observed in the context of an extended, disturbed neutral-gas structure. This paper is structured as follows: In Section~2, we describe the HST/COS spectrum and H\,{\sc i} data. Section~3 presents our main results. Section~$4$ provides further discussion of the redshifted gas component, and Section~$5$ summarizes the main results. Throughout this paper, we adopt the standard $\Lambda$CDM cosmology with $\Omega_{\rm m}=0.3$, $\Omega_{\Lambda}=0.7$, and $H_0=70\ {\rm km\ s^{-1}\ Mpc^{-1}}$. For NGC\,4214, we adopt the heliocentric velocity of $291\pm1\ {\rm km\ s^{-1}}$ as the systemic velocity \citep{Lelli14b}.

\begin{figure*}[t]
    \centering
    \includegraphics[width=1\textwidth]{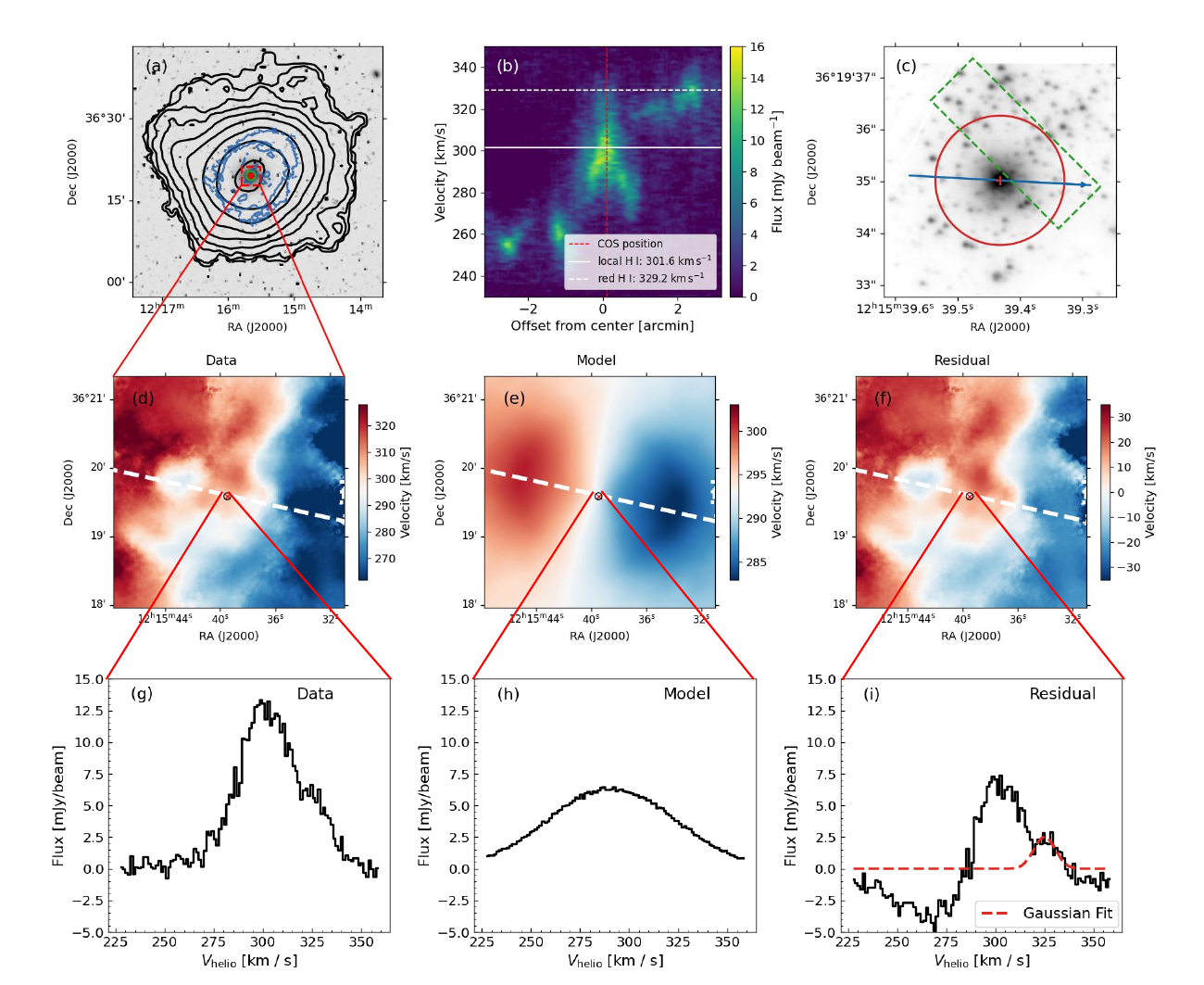}
    \caption{(a): H\,{\sc i} column-density map of NGC\,4214 from the FAST observation shown in black contours overlaid on the DESI--$r$ band image \citep{DESI2026}. The black contours correspond to a column density from $10^{17}\ \rm cm^{-2}$ to $10^{21}\ \rm cm^{-2}$. The blue contours represent the H\,{\sc i} column-density map of the VLA data from THINGS. The column density is $10^{19}\ \rm cm^{-2}$, $10^{20}\ \rm cm^{-2}$, $10^{21}\ \rm cm^{-2}$ and $10^{22}\ \rm cm^{-2}$. The COS line of sight is shown by a red cross in each panel. The green circle marks the HIRES line of sight. The central disk region is shown in a red square, whose size is $3\farcm38\times3\farcm38$. (b): The PV diagram along the major axis (white dashed line in panel (d)--(f)). The red dashed line marks the position of the COS sightline. The white solid and dashed lines mark the local and redshifted H\,{\sc i} velocity, respectively. (c): The HST/ACS-F125LP UV image overlaid with the COS PSA (red circle). The blue arrow shows the direction of increasing wavelength along the COS dispersion axis. The green rectangle represents the $1.148''\times3.5''$ slit at PA $45^\circ$ of HIRES. The green rectangle shows the nominal, header-derived HIRES slit position without any artificial displacement. Its absolute registration relative to the HST image remains uncertain, and subarcsecond offsets are not discernible at the scale of panel (a). (d): The velocity map from VLA--THINGS data in the zoom-in region. The black circle represents the beam size of VLA ($6''$). (e): The rotation disk model fitted by 3dBbarolo. (f): The residual of the rotation disk model. (g)--(i): The extracted H\,{\sc i} spectrum in the COS line of sight. The red dashed line in panel (i) shows the redshifted Gaussian component.}
    \label{fig:h1_cos_desi}
\end{figure*}

\begin{figure*}[htbp]
    \centering
    \includegraphics[width=0.37\textwidth]{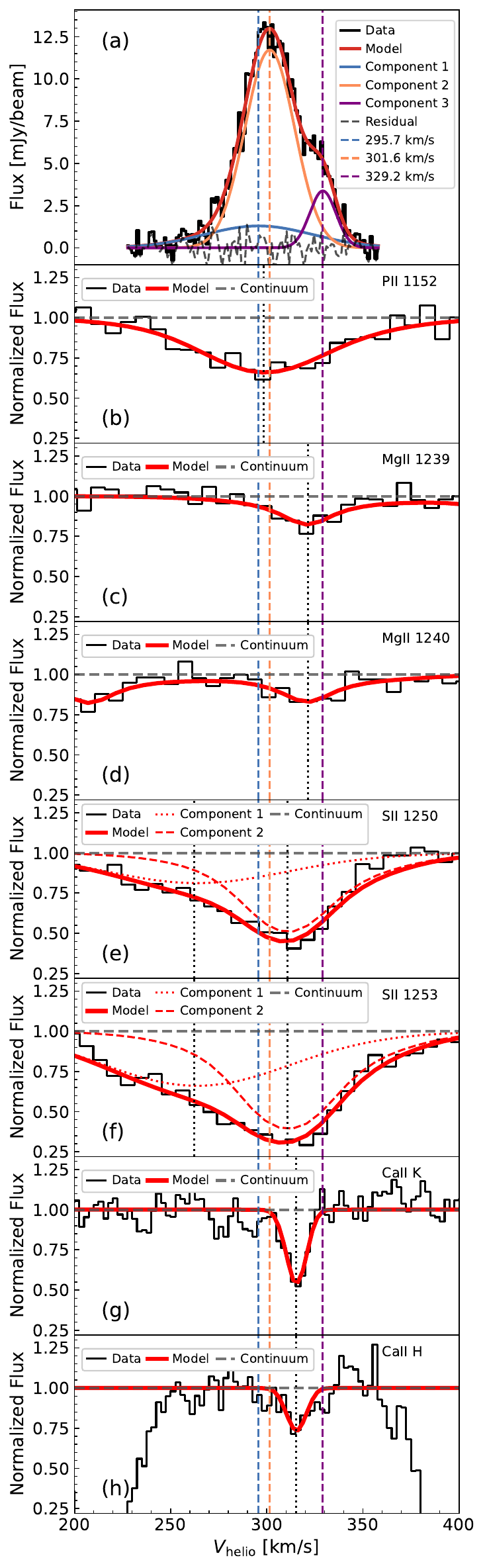}
    \caption{(a): VLA H\,{\sc i} spectrum of the central region with COS observations. The data and the fitted model are shown in black histograms and red lines, respectively. The residuals are represented using a gray dashed line. The blue, orange and purple dashed lines in each panel indicate three different H\,{\sc i} components at 295.7, 301.6, and 329.2 $\rm km\ s^{-1}$, respectively. (b)-(h): Normalized absorption-line profile of NGC\,4214. We plot the observed spectrum using a black histogram and the fitted model using a red solid line in each panel. The gray dashed line in each panel is the continuum. The line center is marked using a black dotted line in each panel. For S\,{\sc ii} $1250$ and $1253$, the two components are represented by red dotted and dashed curves, respectively. The Ca\,{\sc ii} spectra are fitted only $272.6$--$362.6\,\mathrm{km\,s^{-1}}$ for K and $287.6$--$347.6\,\mathrm{km\,s^{-1}}$ for H on the zero-point-corrected heliocentric velocity scale.}
    \label{fig:h1_cos_spec}
\end{figure*}

\section{Method}

\subsection{HST/COS data}

We obtained HST/COS x1d spectrum of G$130$M grating from the Hubble Spectroscopic Legacy Archive \citep[HSLA;][]{Sankrit25}. There are $10$ observations in NGC\,4214 by $4$ programs ($11579$, $15193$, $17106$ and $17465$). For each observation, we estimated the SNR of each absorption feature using a $10$ \AA\ absorption-line-free window. We adopted the spectrum from program $11579$ (PI: Alessandra Aloisi), which provides SNR $>10$ for all elements analyzed in this work. The observations used the COS G130M grating with the 1291 \AA central-wavelength setting, the $2.5''$-diameter Primary Science Aperture, and detector lifetime position LP1. Four TIME-TAG exposures were obtained at FP-POS $=1,2,3$, and $4$ with exposure times of 492.032, 492.032, 492.032, and 492.000 s, respectively, giving a total exposure time of 1968.096 s. Both FUV detector segments were active, providing a nominal wavelength range of approximately 1130--1420 \AA. The remaining COS programs do not provide an independent confirmation of the narrow redshifted component: program 15193 provides an inconclusive lower-S/N comparison, program 17106 lacks complete S\,{\sc ii}/Mg\,{\sc ii} coverage, and program 17465 samples spatially offset sightlines and does not yield robust S\,{\sc ii} or Mg\,{\sc ii} detections. The final spectrum is binned by $3$ pixels to increase the SNR \citep{Jiang26}. The line of sight of this observation is marked by a red cross in Figure \ref{fig:h1_cos_desi}(a). Figure \ref{fig:h1_cos_desi}(c) shows the HST/ACS UV image overlaid on the COS primary science aperture (PSA). The blue arrow shows the direction of increasing wavelength along the COS dispersion axis. \\

To correct the known instrumental shift issue of the COS spectrum \citep{Wakker15}, we used Galactic H\,{\sc i} $21\ \rm cm$ emission as an external velocity reference. We extracted HI4PI H\,{\sc i} spectra within $0.2^{\circ}$ of the COS sightline \citep{HI4PI16,Qu22}. Because the Galactic H\,{\sc i} has multiple velocity components, we adopted the intensity-weighted mean velocity over the relevant Galactic velocity interval as the reference H\,{\sc i} velocity. The Galactic absorption lines of low-ionization-state ions are assumed to have the same velocity centroid as the H\,{\sc i} $21\ \rm cm$ lines. Among the available transitions, we selected Galactic S\,{\sc ii} absorption lines because they are relatively free from saturation and blending with unrelated features. We establish a local wavelength reference using Galactic S\,{\sc ii} absorption and the corresponding Galactic H\,{\sc i} emission. The Galactic S\,{\sc ii} reference and the NGC\,4214 S\,{\sc ii}/Mg\,{\sc ii} features lie in the same local wavelength region, reducing sensitivity to wavelength-calibration errors that vary over larger wavelength scales. We propagate the fitted Galactic S\,{\sc ii} centroid uncertainty into the absorber velocities. A possible physical mismatch between the H\,{\sc i}-emission and S\,{\sc ii}-absorption centroids remains a systematic uncertainty. Comparing the S\,{\sc ii} centroid with the reference H\,{\sc i} velocity yields a COS velocity zero-point offset of $0.02\ \rm km\ s^{-1}$. The correction of $0.02\ \rm km\ s^{-1}$ is much smaller than the statistical velocity uncertainties of the NGC\,4214 absorbers and does not affect the identification of the redshifted components. We then applied this offset to correct the velocities of all COS absorbers. \\

We selected P\,{\sc ii}, S\,{\sc ii}, and Mg\,{\sc ii} because these transitions are sufficiently unblended, have usable continua, and allow the velocity structure of the low-ionization gas to be measured. We also inspected the available C\,{\sc ii}, C\,{\sc ii}*, O\,{\sc i}, Si\,{\sc ii}, Fe\,{\sc ii}, Fe\,{\sc iii}, Al\,{\sc ii}, and N\,{\sc i} transitions, but these lines are saturated, blended, broad, or too weak to isolate a narrow component at the velocity of the redshifted S\,{\sc ii}, and Mg\,{\sc ii} absorption. We show the line details in columns $1$-$3$ of Table \ref{tab:absorbers}. In Table \ref{tab:absorbers}, Column $1$ shows the ions, while column $2$ and $3$ are the wavelength at rest frame and oscillator strength. The references of the line properties are listed in column $9$. The SNR is calculated over a $10$ \AA\ absorption-line-free window for each absorption feature, and it is listed in column $4$ of Table \ref{tab:absorbers}. \\

We then masked absorption features and fitted the continuum with a polynomial. Then we fit absorption signals with a Voigt profile (VP) convolved with the COS--G$130$M line-spread function \footnote{\url{https://spacetelescope.github.io/hst_notebooks/notebooks/COS/LSF/LSF.html}}. We accounted for the spatially extended UV illumination of the COS PSA shown in Figure \ref{fig:h1_cos_desi}(c). We projected the ACS/F125LP surface-brightness distribution onto the on-sky COS dispersion direction and converted it into a wavelength-broadening kernel. The projected surface-brightness profile was referenced to the COS PSA center, so the spatial kernel includes both the illumination-induced centroid shift and the additional broadening along the dispersion direction. These correspond to a redward displacement of approximately $2.1\ \mathrm{km\,s^{-1}}$ before the Galactic zero-point correction. And the combined spatial and instrumental response has a central FWHM of $36.9\ \mathrm{km\,s^{-1}}$. As a sensitivity test, we displaced the assumed aperture center by $\pm0.1''$ along the dispersion direction. We re-calculated both the image-based response and the Galactic S\,{\sc ii} wavelength correction for each displacement. Under the assumption of spatially uniform absorption across the illuminating source, the induced shift is largely shared by the Galactic and NGC\,4214 absorption and therefore largely cancels in the final calibrated velocities. The calibrated Mg\,{\sc ii} and S\,{\sc ii} centroids varied by at most $0.75\ \rm km\,s^{-1}$ and $0.58\ \rm km\,s^{-1}$. The adopted displacement is a test scale rather than a measured registration uncertainty.  The Voigt-profile model was then convolved with both this spatial kernel and the COS point source LSF. We then adopted simultaneous fitting for the doublet absorption lines. Figure \ref{fig:h1_cos_spec}(b)-(f) indicates the normalized spectrum of P\,{\sc ii}, S\,{\sc ii} and Mg\,{\sc ii} absorption lines. The dotted line in each panel shows the center of velocity. The Mg\,{\sc ii} velocity centroid is well constrained and supports the redshifted component, but its column density remains poorly constrained because of parameter degeneracy. The S\,{\sc ii} doublet is fitted with two velocity components. The lower-velocity component has $v_{\rm helio}=262.9\pm24.2\ {\rm km\ s^{-1}}$, which is near the adopted systemic velocity within the uncertainty. The lower velocity S\,{\sc ii} component is not strong evidence for anomalous gas, but it does show that the foreground cool gas is kinematically complex. We therefore do not interpret this component as a clear anomalous gas. Instead, we focus on the redshifted S\,{\sc ii} component at $v_{\rm helio}=310.8\pm2.4\ {\rm km\ s^{-1}}$, because it remains offset from the systemic velocity. The presence of the lower-velocity S\,{\sc ii} component nevertheless indicates that the foreground low-ionization gas is multi-component, so the redshifted component should not be regarded as the only cool-gas structure along this sightline. The heliocentric velocity and column density from VP fitting are also shown in column $5$ and $8$ of Table \ref{tab:absorbers}. 
\\

\subsection{Keck/HIRES data}

We retrieved archival observations of NGC\,4214 obtained with HIRES on the Keck\,I telescope \citep{Vogt94} from the Keck Observatory Archive.\footnote{\url{https://koa.ipac.caltech.edu/}} We selected four exposures obtained on 2000 January 31 UT under program H37bH (PI: W.~Vacca), targeting the central star-forming region designated NGC4214K1 (green circle in Figure~\ref{fig:h1_cos_desi}(a)). These exposures correspond to frames 074--077 and have integration times of 1200~s each, giving a total exposure time of 4800~s. The observations used $1\times1$ detector binning and the C4 decker, with a nominal slit width and length of $1.148^{\prime\prime}\times3.5^{\prime\prime}$ (Figure~\ref{fig:h1_cos_desi}(c)). The header-reported resolving power is $R=35{,}800$, corresponding to a nominal velocity resolution of $c/R\simeq8.4~\mathrm{km\,s^{-1}}$. The usable extracted orders span approximately 3933--6298~\AA, with gaps between some orders. The mean header pointing is approximately $\alpha_{\mathrm{J2000}}=12^{\mathrm h}15^{\mathrm m}39.41^{\mathrm s}$ and $\delta_{\mathrm{J2000}}=+36^\circ19^\prime35.7^{\prime\prime}$. Observations at other pointings were not combined with these four exposures. \\

For each exposure and order, we estimated a local linear continuum and normalized the extracted spectrum. The normalized spectra were resampled onto a common $2.5~\mathrm{km\,s^{-1}}$ grid using native-pixel overlap and combined with fixed weights derived from the continuum noise of each exposure and order. Thorium--argon (ThAr) arc-lamp exposures were used to calibrate the wavelength scale and characterize the instrumental spectral response. Both overlapping orders containing Ca\,{\sc ii} K were used, whereas Ca\,{\sc ii} H was taken from one order. The coadded spectrum has continuum S/N values per resolution element of 21.5 near Ca II K and 19 near Ca II H. We quantify the Ca II absorption-line S/N as $W_r/\sigma(W_r)$, where $W_r$ is the rest-frame equivalent width integrated over $295$--$335{\rm km\,s^{-1}}$ before the velocity zero-point correction. We propagate $\sigma(W_r)$ using the full covariance matrix of the continuum-normalized stacked spectrum. Integrating the continuum-normalized stacked spectra over $295$-$335\,\mathrm{km\,s^{-1}}$ before the velocity zero-point correction gives rest-frame equivalent widths of $W_r(\mathrm{K})=83.9\pm12.6$~m\AA\ and $W_r(\mathrm{H})=68.3\pm18.0$~m\AA, corresponding to $W_r/\sigma(W_r)=6.65$ and $3.79$, respectively, with statistical uncertainties propagated from the full covariance matrix. As mentioned before, NGC\,4214 is a diffuse source. To account for non-uniform slit illumination, we constructed a fiducial source-dependent spectral response using archival HST/WFC3 F336W and F438W images from program 11360. The across-slit illumination was mapped into velocity using a conversion magnitude of $7.10~\mathrm{km\,s^{-1}\,arcsec^{-1}}$. This conversion was obtained by modeling isolated ThAr profiles as a uniformly illuminated slit convolved with a residual Gaussian response and integrated over detector pixels. The corresponding slit velocity span and residual Gaussian dispersion were 8.16 and $1.35~\mathrm{km\,s^{-1}}$, respectively. For the science spectra, the image-derived slit response was convolved with this residual Gaussian, without adding a second full slit-broadened instrumental profile. The absorption model was then passed through the same native-pixel integration, continuum-normalization, and stacking operations as the data. We fitted Ca\,{\sc ii} K and H simultaneously with a single-component Voigt absorption model. \\

We applied a conditional velocity zero-point correction using Galactic Ca\,{\sc ii} absorption and a selected Galactic H\,{\sc i} component from HI4PI \citep{HI4PI16}. We obtained the additive correction of $+2.56~\mathrm{km\,s^{-1}}$. The final fitted results are shown in Table~\ref{tab:absorbers}. The simultaneous Ca\,{\sc ii} K and H fit was restricted to heliocentric velocity intervals of $270$--$360$ and $285$--$345\,\mathrm{km\,s^{-1}}$, respectively, before applying the velocity zero-point correction. Figure~\ref{fig:h1_cos_spec}(g) and (h) present the spectrum of Ca\,{\sc ii} K and H signals. Figure~\ref{fig:h1_cos_spec} displays both spectra and their best-fitting models over $200$--$400\,\mathrm{km\,s^{-1}}$ on the corrected velocity scale for consistency with the COS panels, but pixels outside the fitting intervals were not included in the absorption-profile fit. The results are also presented in Table~\ref{tab:absorbers}. The quoted Ca\,{\sc ii} uncertainties are approximately $1\sigma$ statistical errors conditional on the adopted full-covering illumination model, with the column-density error derived from the local fit covariance. The velocity error includes the statistical uncertainty of the fitted Galactic Ca\,{\sc ii} reference, but uncertainties in the adopted Galactic H\,{\sc i} reference velocity and component association, together with systematic uncertainties from illumination and continuum/background modeling, are not included in the quoted errors.

\subsection{H\textsc{i} Spectrum}

To investigate the H\,{\sc i} content in the central disk, we used the VLA--THINGS data \citep{Walter08}. The column density map from VLA is shown by blue contours in Figure \ref{fig:h1_cos_desi}(a). The optical radius ($0.5D_{25}$) of NGC\,4214 is $3\farcm38$ \citep{Karachentsev04}, we then marked the central stellar disk region using a red square in Figure \ref{fig:h1_cos_desi}(a). Figure \ref{fig:h1_cos_desi}(b) represents the position-velocity (PV) diagram of the central disk along the major axis of VLA data, with PA $=65^\circ$ and slice width of $6''$ (white dashed line in panel (d)--(f)). The location of COS sightline is marked by a red dashed line. The PV diagram shows high-velocity ($>300\ \rm km\ s^{-1}$) H\,{\sc i} component near the COS position, but this emission extends over $\sim3'$ along the major axis. Figure \ref{fig:h1_cos_desi}(d) represents the H\,{\sc i} velocity map of the central disk. We fitted this region with a regular rotating-disk model using 3dBbarolo \citep{DiTeodoro15}, as shown in Figure~\ref{fig:h1_cos_desi}(e). The residual velocity field in Figure~\ref{fig:h1_cos_desi}(f) reveals remaining velocity structures after subtracting the model, suggesting that the central H\,{\sc i} kinematics include non-circular motions in addition to regular disk rotation. \\

We extracted an H\,{\sc i} spectrum at the COS line-of-sight position to characterize the neutral gas associated with the UV absorption sightline. Since the beam size of VLA observation in THINGS is $6''$, we then extracted the spectrum in the COS LOS region in one VLA beam (black circle in Figure \ref{fig:h1_cos_desi}(d)--(f)). The extracted spectrum from the data, model, and residual is shown in Figure \ref{fig:h1_cos_desi} (g)--(i). At the COS position, the rotating disk model predicts $v_{\rm mod}=290.94\pm0.01\ {\rm km\ s^{-1}}$. The red H\,{\sc i} component at $329.2\pm0.9\ {\rm km\ s^{-1}}$ therefore exceeds the model velocity by $38.3\pm0.9\ {\rm km\ s^{-1}}$, producing the redshifted residual shown in Figure \ref{fig:h1_cos_desi} (i). The fitted spectrum is shown in Figure \ref{fig:h1_cos_spec}(a). We compared two- and three-Gaussian models for the VLA H\,{\sc i} spectrum. For $102$ fitted channels in the spectrum, the fitted result improves from $\chi^2 = 193.83$ for the two-Gaussian model to $\chi^2 = 160.01$ for the three-Gaussian model, corresponding to reduced $\chi^2$ values of $2.02$ and $1.72$, respectively. Information-criterion tests also show better results with the three-Gaussian model, with a lower BIC by $19.9$. We therefore adopt the three-Gaussian decomposition to fit the asymmetric H\,{\sc i} profile. The properties of three components are presented in Table \ref{tab:ngc_vla_h1}. We calculated the column density using the formula \citep{Meyer17}:
\begin{equation}
    (\frac{N_{\mathrm{HI}}}{\mathrm{cm^{-2}}})=1.25\times10^{21}\times (\frac{S_{\mathrm{int}}}{\mathrm{mJy\ km\ s^{-1}}})\times(\frac{\Omega_{\mathrm{beam}}}{\mathrm{arcsec^{2}}})^{-1},
\end{equation}
where $S_{\mathrm{int}}$ is the integrated flux of each Gaussian component, and $\Omega_{\mathrm{beam}}$ is the beam area of VLA. The final model is represented using a red curve in Figure \ref{fig:h1_cos_spec}(a). The three components are marked using orange, blue, and purple curves, respectively. We use the dashed line to mark the central velocity of each component. The main spectral component has a velocity of $301.6\pm0.4\ {\rm km\ s^{-1}}$. We treat it as the local H\,{\sc i} component. The highest-velocity component is centered at $329.2\pm0.9\ {\rm km\ s^{-1}}$. We consider it as the redshifted H\,{\sc i} component, and its column density is $\log (N_{\rm HI}/{\rm cm^{-2}})=21.25\pm0.17$. \\

The extended H\,{\sc i} data of NGC\,4214 were obtained from FASHI survey. The source was identified using the H\,{\sc i} Source Finding Application (SoFiA) version $2$ \citep{Serra15,Westmeier21,Westmeier22}. We adopted a $6.5\sigma$ threshold to better separate NGC\,4214 from nearby dwarf galaxies. The column density map is represented using black contours in Figure \ref{fig:h1_cos_desi}(a). The heliocentric velocity measured from the FAST H~{\sc i} profile is $287.96\pm0.01\ {\rm km\ s^{-1}}$. This value is close to the adopted systemic velocity from NED, and the small difference does not affect the identification of the redshifted components discussed below. The line widths measured at $20\%$ and $50\%$ of the peak flux are $W_{20}=86.08\pm0.02\ {\rm km\ s^{-1}}$ and $W_{50}=61.44\pm0.01\ {\rm km\ s^{-1}}$, respectively.  We calculated the uncertainty of the velocity using $\sigma_{\mathrm{vel}}=3\sqrt{P\delta V}/\rm SNR$ \citep{Fouque90,Koribalski04}, where SNR$=5710.88$ for this source. $P=(W_{20}-W_{50})/2$ is a measure of the steepness of the profile edges. $\delta V$ is the spectral resolution, which is $6.4\ \rm km\ s^{-1}$. The uncertainty of $W_{20}$ and $W_{50}$ is estimated by $3\sigma_{\mathrm{vel}}$ and $2\sigma_{\mathrm{vel}}$. We then calculated H\,{\sc i} mass via the formula \citep{Meyer04,Ellison18}:
\begin{equation}
\frac{M_{\mathrm{HI}}}{M_\odot}=\frac{2.356\times10^{5}}{1+z}(\frac{D}{\mathrm{Mpc}})^{2}\frac{S_{\mathrm{int}}}{\mathrm{Jy\ km\ s^{-1}}},
\label{eq:h1mass}
\end{equation}
where $D$ is the distance of NGC\,4214. We calculated $S_{\mathrm{int}}$ by integrating the flux in all channels in the moment0 map recognized by SoFiA. The total H\,{\sc i} mass is $\sim7.08\times10^8\ M_\odot$. NGC~\,4214 has an H\,{\sc i}-to-stellar mass ratio of $M_{\rm HI}/M_*\sim0.47$. Using the Local Volume catalog from \citet{Karachentsev13}, we calculated the median H\,{\sc i}-to-stellar mass ratio of $0.29$, with a 16th--84th percentile range of 0.17--0.52 for galaxies with $9<\log(M_*/M_\odot)<9.5$. The value for NGC~\,4214 lies within this calculated range, indicating that the galaxy is not unusually H\,{\sc i} poor for its stellar mass.  \\


\subsection{Extended H\textsc{i} Kinematics}
To describe the extended H\,{\sc i} kinematics, we modeled the velocity field from FASHI with {\sc 3dBbarolo} \citep{DiTeodoro15}, allowing the position angle and inclination to vary with radius. We used a tilted ring with a ring width $1.22'$. The results are shown in Figure \ref{fig:h1_3d_model}. The residual map was constructed by subtracting the best-fit model velocity field from the observed velocity field. For visualization, we separate the disk into inner and outer regions based on the fitted kinematic structure \citep{Cao22}. In Figure \ref{fig:h1_3d_model}, the red dashed circle shows the outer boundary of the inner disk, while the black dashed circle shows the inner boundary of the outer disk. The solid lines in Figure \ref{fig:h1_3d_model}(a) mark the average position angle of the inner and outer disks, with values of $77.3^\circ$ and $136.6^\circ$, respectively. \\

\section{Results}

\begin{deluxetable}{ccccccccc}
\scriptsize
\setlength{\tabcolsep}{0.6mm}
\tablecaption{Properties of Absorption Lines}
\label{tab:absorbers}
\tablewidth{0pt}
\tablehead{
\colhead{Ion} & \colhead{$\lambda_{\mathrm{res}}$} & \colhead{$f$} & \colhead{SNR} & \colhead{Heliocentric Velocity} & \colhead{$\Delta v_{\rm sys}$} & \colhead{$\Delta v_{\rm loc}$} & \colhead{$\log N$(fit)} & \colhead{Ref}\\
 & \colhead{(\AA)} & \colhead{($\times10^{-3}$)} & & \colhead{($\mathrm{km\ s^{-1}}$)} & \colhead{($\mathrm{km\ s^{-1}}$)} & \colhead{($\mathrm{km\ s^{-1}}$)} & \colhead{($\mathrm{cm^{-2}}$)}
}
\colnumbers
\startdata
P\,{\sc ii} & 1152.818 & 256 & 41.5 & $298.5\pm3.3$ & $7.5\pm3.4$ & $-3.1\pm3.3$ & $13.72\pm0.03$ & \citetalias{Cashman24} \\
Mg\,{\sc ii} & 1239.925, 1240.395 & 0.621, 0.351 & 44.8 & $321.3\pm2.8$ & $30.3\pm3.0$ & $19.7\pm2.8$ & unconstrained & NIST-ASD \\
S\,{\sc ii} (comp1) & 1250.584, 1253.811 & 6.02, 12.1 & 53.7 & $262.9\pm24.2$ & - & - & $15.10\pm0.22$ & NIST-ASD \\
S\,{\sc ii} (comp2) & - & - & - & $310.8\pm2.4$ & $19.8\pm2.6$ & $9.2\pm2.4$ & $15.54\pm0.11$ & NIST-ASD \\
Ca\,{\sc ii} (K, H) & 3934.775, 3969.590 & 626.7, 311.6 & 21.5, 19.0 & $316.6\pm1.1$ & $25.6\pm1.5$ & $15.0\pm1.2$ & $12.07\pm0.05$ & \citetalias{Murphy14} \\
\enddata
\tablecomments{Column (1): Ions used in this work. Column (2): The line wavelength in rest frame. Column (3): Oscillator strength. Column (4): Continuum SNR per resolution element. Column (5): Heliocentric velocity from VP fitting. Column (6): The velocity offset to the systemic velocity of NGC\, 4214. Column (7): The velocity offset to the local velocity of NGC\, 4214 from VLA. Column (8): Column density from VP fitting. Column (9): The references of line properties ($\lambda_{\mathrm{res}}$ and $f$). The atomic data for P\,{\sc ii} are from \citet{Cashman24}, those for Mg\,{\sc ii} and S\,{\sc ii} are from NIST-ASD \citep{NIST_ASD}, and those for Ca\,{\sc ii} are from \citet{Murphy14}. For S\,{\sc ii}, only component $2$ is used to quantify the redshifted absorber; component $1$ represents the lower-velocity absorption component. }
\end{deluxetable}

\begin{deluxetable}{ccccccc}
\setlength{\tabcolsep}{3mm}
\tablecaption{Properties of Fitted Gaussian Components in VLA Spectrum}
\label{tab:ngc_vla_h1}
\tablewidth{0pt}
\tablehead{
\colhead{Component} & \colhead{Peak Flux} & \colhead{Central Velocity} & \colhead {$\sigma_V$} & \colhead{$\Delta v_{\rm sys}$} & \colhead{$\Delta v_{\rm loc}$} & \colhead{$\log N$} \\
\colhead{(No.)} & \colhead{($\rm mJy\ beam^{-1}$)} & \colhead{($\rm km\ s^{-1}$)}  & \colhead{($\rm km\ s^{-1}$)} & \colhead{($\rm km\ s^{-1}$)} & \colhead{($\rm km\ s^{-1}$)} & \colhead{($\rm cm^{-2}$)}
}
\colnumbers
\startdata
$1$ & $1.3\pm1.1$ & $295.7\pm10.1$ & $28.2\pm8.7$ & $4.7\pm10.1$ & $-5.9\pm10.1$ & $21.45\pm0.27$ \\
$2$ & $11.7\pm1.1$ & $301.6\pm0.4$ & $12.9\pm0.8$ & $10.6\pm1.2$ & - & $22.07\pm0.06$ \\
$3$ & $3.4\pm0.5$ & $329.2\pm0.9$ & $6.8\pm1$ & $38.2\pm1.3$ & $27.6\pm1.0$  & $21.25\pm0.17$
\enddata
\tablecomments{Column (1): Component number. Column (2)--(4): Peak flux, centroid velocity, and Gaussian velocity width ($\sigma$) of each Gaussian component. Column (5): The velocity offset relative to the systemic velocity of NGC\,4214. Column (6): The velocity offset to the local velocity. Column (7): Column density of each component. The errors of the Gaussian components are obtained from the covariance matrix. The errors of the velocity offset and column density are calculated via error propagation.}
\end{deluxetable}

We further calculated the velocity offsets of the UV and optical absorbers relative to both the systemic velocity and the local H\,{\sc i} velocity of NGC\,4214. The results are listed in Columns $6$ and $7$ of Table~\ref{tab:absorbers}. The Ca\,{\sc ii} absorber has a redshifted velocity relative to the systemic velocity by $\sim26\ \rm km\ s^{-1}$. The Mg\,{\sc ii} and S\,{\sc ii} absorption centroids are redshifted to the systemic component by $\sim30$ and $\sim20\ {\rm km\ s^{-1}}$, respectively. These offsets exceed their fitted centroid uncertainties and persist after applying the image-based spatial-illumination correction. The redshifted velocity of P\,{\sc ii} remains tentative because its offset, $\Delta v_{\rm sys}=7.5\pm3.4\ \rm km\ s^{-1}$, has only approximately $2.2\sigma$ significance. \\

Comparing with the local H\,{\sc i} component, both the UV/optical absorption lines and the H\,{\sc i} spectrum indicate the presence of redshifted gas components toward the central region of NGC\,4214. The S\,{\sc ii}, Mg\,{\sc ii} and Ca\,{\sc ii} absorption lines are detected, but their velocity significance differs. The redshifted S\,{\sc ii} component provides the strongest UV evidence for redshifted gas, while Mg\,{\sc ii} independently supports a redshifted centroid. However, the Mg\,{\sc ii} column density remains poorly constrained. Their centroids are redshifted relative to both the systemic velocity and the local H\,{\sc i} component, although the local-frame offsets are more modest than the systemic-frame offsets. The PV diagram in Figure \ref{fig:h1_cos_desi}(b) also shows that the high-velocity H\,{\sc i} is extended along the major axis rather than confined to a compact isolated cloud, suggesting that it may be related to warped or non-circular disk motions. Relative to the dominant local H\,{\sc i} component ($301.6\pm0.4\ \rm km\ s^{-1}$), the redshifted H\,{\sc i} component remains offset by $\sim28\ \rm km\ s^{-1}$, indicating non-circular central gas kinematics. Together with the redshifted VLA H\,{\sc i} component, the UV absorbers indicate anomalous redshifted cool gas toward the central region of NGC\,4214. This component may be associated with the disturbed central gas kinematics, although the present data do not establish a direct causal connection. The UV absorption profiles appear broader than the local H\,{\sc i} emission component. Part of the broadening may arise from the combined effects of the COS point-source LSF and the spatially extended illumination of the PSA, while unresolved velocity structure may also contribute. The intrinsic absorption-line width therefore remains poorly constrained. Differences in gas phase and spatial weighting may also contribute to the line-width difference. The 21-cm emission traces neutral gas, whereas S\,{\sc ii}, Mg\,{\sc ii} and Ca\,{\sc ii} can arise in both neutral and partially ionized material. \\

\begin{figure*}[htbp]
    \centering
    \includegraphics[width=1\textwidth]{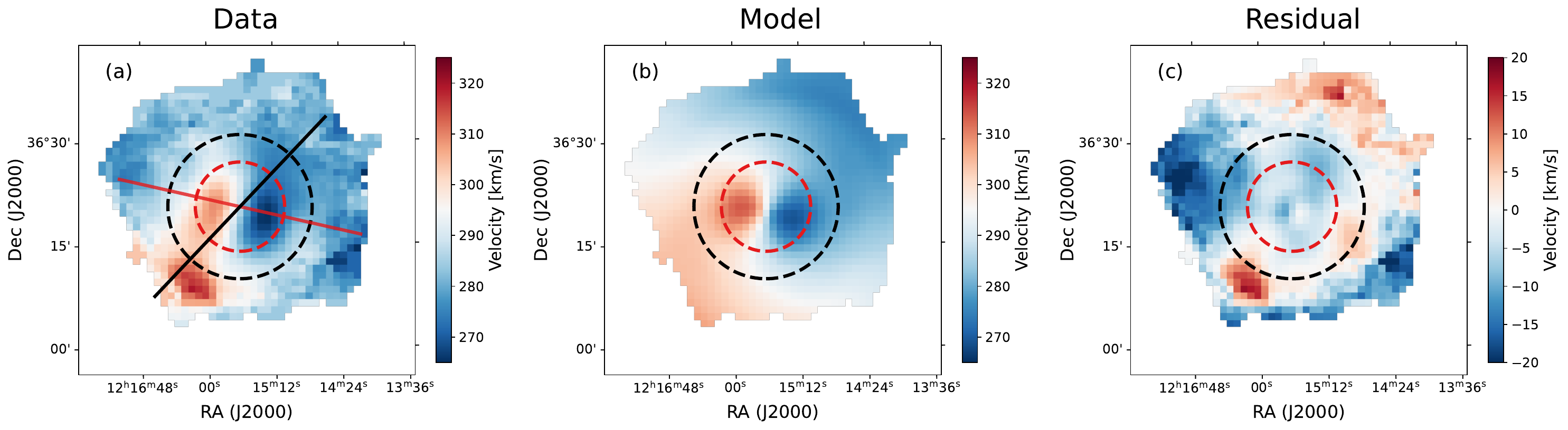}
    \caption{$3$D warped disk model of H\,{\sc i} data. The data, model, and residual are shown in (a), (b), and (c). The color map represents the velocity level. In each panel, the red dashed circle shows the outer boundary of the inner disk, while the black dashed circle shows the inner boundary of the outer disk. The position angle of the inner ($77.3^{\circ}$) and outer ($136.6^\circ$) disk is marked by a red and a black line in panel (a). The red and black dashed circles have radii of $5.5'$ and $9.5'$, respectively.}
    \label{fig:h1_3d_model}
\end{figure*}

Figure~\ref{fig:h1_3d_model} shows that the extended H\,{\sc i} disk is warped or kinematically twisted. We do not interpret the FAST residuals as direct evidence for the origin of the COS absorber. Instead, the FAST data provide the large-scale neutral-gas context, showing that the central redshifted component is embedded in an extended, disturbed H\,{\sc i} reservoir. NGC 4214 has a relatively low inclination of $i\sim30^\circ$ \citep{Lelli14b}. In this geometry, vertical motions such as fallback onto the disk can contribute to the observed line-of-sight velocity, whereas purely in-plane streaming motions are geometrically diluted. Thus, the foreground redshifted absorption is consistent with fallback or inflow, although non-circular disk motions and projection effects remain viable alternatives. Figure~\ref{fig:h1_3d_model}(c) also reveals a redshifted residual component in the outer disk (higher than $10\ \rm km\ s^{-1}$), indicating that the extended H\,{\sc i} distribution is also kinematically disturbed. Such residual structure is not fully reproduced by the present warped-disk model. \\

\section{Discussion}

NGC\,4214 is a Wolf-Rayet galaxy hosting a central stellar population with an age of 4-5 Myr \citep{Leitherer96,Drozdovsky02}, indicating a recent, intense starburst episode. The central region also has numerous Wolf-Rayet stars \citep{Sargent91,MasHesse99,Drozdovsky02,Srivastava11}. The active star formation in the galaxy can enrich the ISM with heavy elements (e.g., Mg, P, and S) through the explosions of massive stars \citep{Iwamoto99,Johnson19}. \citet{Sanchez23} reported that NGC\,4214 has superbubbles and supernova remnants, with the expansion velocity up to $\sim80\ \rm km\ s^{-1}$. As a result, the metals may be ejected into the CGM through the supernova-driven feedback. Over long timescales, this metal-enriched CGM may be recycled back into the ISM during the galaxy's evolution \citep{Fraternali08,Marinacci10,Fraternali17}. Because absorption can only be produced by gas located in front of the UV continuum source, the redshifted low-ionization absorption has a direct geometrical implication. Since the absorbers lie in the foreground of the central UV source, their positive velocity offsets indicate motion away from the observer and toward the background UV source in projection. This geometry is consistent with fallback or inflow and disfavors a simple nearside expanding outflow, although warped-disk motions, non-circular motions, and projection effects remain viable alternatives. Therefore, the redshifted Mg\,{\sc ii} and S\,{\sc ii} absorption is naturally interpreted as candidate inflowing or fallback cool gas. The redshifted low-ionization absorbers and the redshifted VLA H\,{\sc i} component are consistent with anomalous cool gas kinematics toward the central region of NGC\,4214. \\

The HIRES spectra reveal Ca\,{\sc ii} absorption at $v_{\rm helio}=316.6\pm1.1\,\mathrm{km\,s^{-1}}$, with $\log[N(\mathrm{Ca\,II})/\mathrm{cm}^{-2}]=12.07\pm0.05$ under the adopted illumination and foreground-calibration assumptions. This component lies approximately $15\,\mathrm{km\,s^{-1}}$ redward of the local H\,{\sc i} component and between the redshifted S\,{\sc ii} and Mg\,{\sc ii} centroids, providing additional evidence for redshifted absorbing material toward the nuclear region. However, differences in aperture sampling and continuum weighting prevent a unique association with the same gas traced by COS. If Ca\,{\sc ii} and the redshifted S\,{\sc ii} component arise in the same gas, their fitted columns imply $\log[N(\mathrm{Ca\,II})/N(\mathrm{S\,II})]=-3.47\pm0.12$. This ionic ratio depends on both differential dust depletion and ionization, since Ca\,{\sc ii} can be photoionized to Ca\,{\sc iii} by photons below the H\,{\sc i} ionization threshold \citep{Guber16}. Adopting the solar abundances of \citet{Asplund09}, $\log({\rm Ca/S})_\odot=-0.78$. If the Ca\,{\sc ii} and redshifted S\,{\sc ii} absorption trace the same gas, their measured column-density ratio is 2.69 dex below the solar elemental Ca/S ratio. Assuming solar intrinsic Ca/S, this difference requires substantial suppression through ionization and/or differential dust depletion. In the absence of differential depletion, the column-averaged gas-phase ion fractions would need to satisfy $f_{\rm CaII}/f_{\rm SII}\simeq2.0\times10^{-3}$. This conditional requirement has not been tested with an ionization model and therefore does not establish that ionization alone can reproduce the observed ratio. Furthermore, the poorly constrained Mg\,{\sc ii} column density, the uncertain association of P\,{\sc ii} with the redshifted component, and the different spatial weighting of HIRES and COS limit a joint interpretation of the ionic columns. The Ca\,{\sc ii} measurement thus constrains the combined effects of ionization and differential depletion under the common-gas assumption, but does not uniquely determine either effect. \\

We further estimated the mass inflow rate ($\dot{M}_{\rm in}$) using $\dot{M}_{\rm in}=M_{\rm in}/t_{\rm in}$, where $M_{{\rm in}}=1.4m_{{\rm{H}}}C\langle N_{{\rm H}}\rangle\pi R_\perp^2$ \citep{Rubin12,Fox19}. We assume that the measured redshifted H\,{\sc i} column is representative of a projected area $C\pi R_\perp^2$, and adopt $t_{\rm in}=d_{\rm in}/v_{\rm in}$. We adopted $\langle N_{{\rm H}}\rangle\sim\langle N_{{\rm HI}}\rangle=10^{21.25}\ \rm cm^{-2}$ and $v_{\rm in}=27.6\ \rm km\ s^{-1}$ from VLA data. For this fiducial estimate, we assumed an areal covering fraction of $C=1$ and $R_\perp=d_{\rm in}$. \citet{Johnson17,Mishra24} investigated low and intermediate ionization absorbers around dwarf galaxies and found that low- and intermediate-ion absorption is relatively rare around dwarfs, whereas O\,{\sc vi} is commonly detected within the virial radius. Though we cannot determine the three-dimensional distance of absorbers from the center of NGC~\,4214, the redshifted component might be located near the central region. The redshifted H\,{\sc i} emission near the COS position extends over approximately $3'$ along the major axis, corresponding to a projected extent of approximately $2.6\ \rm kpc$ or a half-extent of $1.3\ \rm kpc$. Given the uncertainty in associating all of this emission with the COS absorber, we then adopted $R_\perp = d_{\rm in} =1\ \rm kpc$. Under these assumptions, we obtain a fiducial mass inflow rate $\dot{M}_{\rm in}\sim1.7\ M_{\odot} {\rm yr^{-1}}$. This fiducial value is approximately 13 times the current SFR of $0.13\ M_\odot\rm yr^{-1}$, but it should be regarded only as an order-of-magnitude estimate because the covering fraction, absorber distance, geometry, and association between the H\,{\sc i} and COS components remain uncertain.  \\

Following \citet{Fox16}, we estimated the gas-phase sulfur abundance using the S\,{\sc ii} and H\,{\sc i} column densities: $[\mathrm{S}/\mathrm{H}]_{\rm app}=\log N(\mathrm{S\,II})-\log N(\mathrm{H\,I})-\log(\rm S/H)_{\odot}$. We adopted $\log(\rm S/H)_{\odot}=-4.88$ \citep{Asplund09}. This gives an apparent sulfur abundance of $[\mathrm{S}/\mathrm{H}]_{\rm app}= -0.83\pm0.20$, where the quoted uncertainty includes statistical column-density errors only. This estimate adopts the redshifted Gaussian H\,{\sc i} component as the associated neutral gas. Using the total one-beam H\,{\sc i} column would further lower the inferred abundance. The nebular abundance of NGC\,4214 is $12+\log(\rm O/H)\sim8.34$ \citep{Kobulnicky96,Walter08}, indicating that the host ISM has roughly $[\rm\alpha/H]\sim-0.35$. Taken at face value, if the adopted redshifted H\,{\sc i} column is physically associated with the S\,{\sc ii} absorber, the gas appears less enriched than the central ISM of NGC\,4214. \citet{Zahedy21} found $\rm\alpha$ element abundances between approximately $-1.3$ and $-0.6$ for a Lyman-limit system associated with a dwarf-galaxy pair at $z\sim0.36$. \citet{Johnson26} measured an oxygen abundance of approximately $-0.6$ in the CGM of an isolated dwarf galaxy at $z\sim0.57$. For the nearby galaxy Sextans B, \citet{Fox26} derived the silicon and carbon abundances of $[\rm Si/H]\sim-1.7$ and $[\rm C/H]\sim-2.1$, respectively, along an inner CGM sightline. These comparisons are consistent with metal-poor external accretion, but they do not uniquely favor this interpretation because the absorber location, ionization correction, pencil-beam/beam-averaged comparison, and correspondence between the UV and H\,{\sc i} components remain uncertain.  \\

Figure \ref{fig:h1_3d_model}(c) indicates that the outer disk exhibits more complex kinematics than the inner disk and is not fully reproduced by the warped-disk model. The large-scale perturbations, such as warps, can induce the asymmetric velocity fields in the disk galaxies \citep{Visser80}. The inflowing gas may drive material onto deeply plunging orbits, thereby affecting the initial radial orbits \citep{Binney92}. The central region of NGC\,4214 can be described by a warped disk model, but the current data do not uniquely determine the physical origin of the warp. In contrast, the outer region exhibits a more complicated velocity field than a warped disk. \citet{Sylos25} reported that a warp alone cannot fully explain all variations in the velocity field. The local perturbations may also contribute to the observed kinematic complexity. External gas perturbations may lead to the asymmetry of the outer velocity field, but the present data do not uniquely identify their origin. Moreover, the extended H\,{\sc i} has a low column density of only $\sim 10^{17}\ \mathrm{cm^{-2}}$, indicating that the outer diffuse gas and the denser central component have different spatial and column-density scales. The current data do not uniquely determine the physical origin of the disturbed gas. The observed kinematic asymmetry may be related to a warp, non-circular motions, or external gas perturbations, but the present analysis does not distinguish among these possibilities. NGC\,4214 was the only dwarf galaxy examined in the present analysis. It was not identified by screening a parent sample for redshifted absorption. Instead, the redshifted absorption was identified through a detailed analysis of archival COS observations selected for their wavelength coverage and signal-to-noise ratio. Consequently, our observations do not provide a formal inflow detection rate. The potential inflows have also been reported in individual dwarfs using UV absorption or spatially resolved H\,{\sc i} and molecular-gas observations, including NGC\,5253 \citep{Lopez12} and UGCA\,105 \citep{Schmidt14}. \citet{Schwartz04} examined six nearby dwarf galaxies using down-the-barrel absorption spectroscopy and did not report a comparable redshifted inflow signature. Therefore, redshifted absorption resembling that detected toward NGC\,4214 appears uncommon in existing nearby-dwarf observations, but its true incidence remains unconstrained by a homogeneous survey.

\section{Summary}

In this paper, we investigated the redshifted cool gas in NGC\,4214. The HST/COS spectrum toward the central star clusters reveals redshifted Mg\,{\sc ii} and S\,{\sc ii} absorption, while P\,{\sc ii} remains tentative. Since the absorbing gas must lie in the foreground of the UV source, its positive velocity offset indicates motion away from the observer and therefore toward the background UV source in projection. The VLA H\,{\sc i} spectrum extracted at the COS position within one VLA beam also shows a redshifted component. We further analyzed the H\,{\sc i} disk using a 3D disk model by 3dBbarolo. The main conclusions are summarized as follows.\\

1. We fitted the UV absorption lines using Voigt profiles. The fitted column densities of P\,{\sc ii} and S\,{\sc ii} are $\log (N/\rm{cm^{-2}})=13.72\pm0.03$, and $15.54\pm0.11$, respectively. The corresponding velocity offsets of P\,{\sc ii}, Mg\,{\sc ii}, and S\,{\sc ii} relative to the systemic velocity of NGC\,4214 are $7.5\pm3.4\ \rm km\ s^{-1}$, $30.3\pm3.0\ \rm km\ s^{-1}$, and $19.8\pm2.6\ \rm km\ s^{-1}$, respectively. The Mg\,{\sc ii} column density is poorly constrained, while the P\,{\sc ii} velocity remains tentative. The Keck/HIRES spectra also reveal a Ca\,{\sc ii} absorption component redshifted by $25.6\pm1.5\ \rm km\ s^{-1}$ relative to the systemic velocity of NGC\,4214. \\

2. The VLA data show that the central H\,{\sc i} disk includes non-circular motion and warped structure. We then extracted the VLA H\,{\sc i} spectrum at the COS sightline, and fitted the H\,{\sc i} emission line using three Gaussian components. The local H\,{\sc i} component has a velocity of $301.6\pm0.4\ {\rm km\,s^{-1}}$, while the redshifted component is centered at $329.2\pm0.9\ {\rm km\,s^{-1}}$. The redshifted component is offset from the systemic velocity of NGC\,4214 by $38.2\pm1.3\ {\rm km\,s^{-1}}$. Relative to the local H\,{\sc i} component, it remains redshifted by $27.6\pm1.0\ {\rm km\,s^{-1}}$. The Mg\,{\sc ii} and S\,{\sc ii} absorbers also show positive velocity offsets relative to the local component, with $\Delta v_{\rm loc}=19.7\pm2.8\ {\rm km\,s^{-1}}$ and $9.2\pm2.4\ {\rm km\,s^{-1}}$, respectively. Under the candidate-inflow interpretation, the fiducial mass inflow rate is $\sim1.7\ M_\odot\rm yr^{-1}$. \\

3. The full H\,{\sc i} disk shows disturbed large-scale kinematics that can be approximately described by a warped or kinematically twisted disk model. The inner region is consistent with a warped disk geometry, while the outer disk shows additional asymmetric residual features. These results indicate disturbed large-scale H\,{\sc i} kinematics, but they do not uniquely constrain the origin of the outer gas. \\

In conclusion, we combined the multiwavelength data to investigate the disturbed gas content of NGC\,4214 on different spatial scales. The redshifted HST/COS UV absorption lines, Keck/HIRES optical signals and the VLA H\,{\sc i} component indicate anomalous cool gas kinematics toward the central region of NGC\,4214. Together with the twisted H\,{\sc i} disk from FAST, these observations show that the anomalous central cool gas occurs in a galaxy with disturbed large-scale H\,{\sc i} kinematics. Future higher-resolution H\,{\sc i} observations, together with molecular-gas and ionized-gas mapping, will be needed to determine whether the redshifted cool gas is inflowing, recycled, or associated with non-circular motions in the disturbed disk. A full understanding of the baryon cycle in dwarf galaxies is essential for constraining galaxy formation and evolution in the early universe. \\

We thank Di Teodoro, Xiao Cao, and Siyue Yu for helpful discussions. This work is supported by the National SKA Program of China No. 2025SKA0150103, National Natural Science Foundation of China under Nos. 12550002, 12133008, 12221003, 12333002, 11890692. We acknowledge the science research grants from the China Manned Space Project with No. CMS-CSST-2021-A04 and No. CMS-CSST-2025-A10. XJ acknowledges the support from grant No. JAT241087, provided by the Fujian Provincial Department of Education. QY was supported by the European Research Council (ERC) under grant agreement No. 101040751. This work uses the data from the FASHI project. FASHI made use of the data from FAST (Five-hundred-meter Aperture Spherical radio Telescope).  FAST is a Chinese national mega-science facility, operated by National Astronomical Observatories, Chinese Academy of Sciences. This research has made use of the HSLA database, developed and maintained at STScI, Baltimore, USA. All the {\it HST} data used in this paper can be found in MAST: \dataset[10.17909/v899-jn69]{http://dx.doi.org/10.17909/v899-jn69}.

\bibliography{references}{}

@ARTICLE{Zhang26,
       author = {{Zhang}, Chuan-Peng and {Zhu}, Ming and {Jiang}, Peng and {Guo}, Hong and {Xu}, Jin-Long and {Liu}, Xiao-Lan and {Yu}, Nai-Ping and {Cheng}, Cheng and {Wang}, Jing and {Wang}, Jie and {FAST Collaboration}},
        title = "{The FAST all sky HI survey DR2: The FASHI catalog and the HI mass function}",
      journal = {Science China Physics, Mechanics, and Astronomy},
         year = 2026,
        month = sep,
       volume = {69},
       number = {12},
          eid = {129811},
        pages = {129811},
          doi = {10.1007/s11433-026-3072-1},
archivePrefix = {arXiv},
       eprint = {2606.31539},
 primaryClass = {astro-ph.GA},
       adsurl = {https://ui.adsabs.harvard.edu/abs/2026SCPMA..6929811Z}
}

@ARTICLE{Guber16,
       author = {{Guber}, C.~R. and {Richter}, P.},
        title = "{Dust depletion of Ca and Ti in QSO absorption-line systems}",
      journal = {\aap},
         year = 2016,
        month = jun,
       volume = {591},
          eid = {A137},
        pages = {A137},
          doi = {10.1051/0004-6361/201628466},
archivePrefix = {arXiv},
       eprint = {1605.03054},
 primaryClass = {astro-ph.GA},
       adsurl = {https://ui.adsabs.harvard.edu/abs/2016A&A...591A.137G}
}

@INPROCEEDINGS{Vogt94,
       author = {{Vogt}, S.~S. and {Allen}, S.~L. and {Bigelow}, B.~C. and {Bresee}, L. and {Brown}, B. and {Cantrall}, T. and {Conrad}, A. and {Couture}, M. and {Delaney}, C. and {Epps}, H.~W. and {Hilyard}, D. and {Hilyard}, D.~F. and {Horn}, E. and {Jern}, N. and {Kanto}, D. and {Keane}, M.~J. and {Kibrick}, R.~I. and {Lewis}, J.~W. and {Osborne}, J. and {Pardeilhan}, G.~H. and {Pfister}, T. and {Ricketts}, T. and {Robinson}, L.~B. and {Stover}, R.~J. and {Tucker}, D. and {Ward}, J. and {Wei}, M.~Z.},
        title = "{HIRES: the high-resolution echelle spectrometer on the Keck 10-m Telescope}",
    booktitle = {Instrumentation in Astronomy VIII},
         year = 1994,
       editor = {{Crawford}, David L. and {Craine}, Eric R.},
       series = {Society of Photo-Optical Instrumentation Engineers (SPIE) Conference Series},
       volume = {2198},
        month = jun,
        pages = {362},
          doi = {10.1117/12.176725},
       adsurl = {https://ui.adsabs.harvard.edu/abs/1994SPIE.2198..362V}
}

@ARTICLE{Murphy14,
       author = {{Murphy}, Michael T. and {Berengut}, Julian C.},
        title = "{Laboratory atomic transition data for precise optical quasar absorption spectroscopy}",
      journal = {\mnras},
         year = 2014,
        month = feb,
       volume = {438},
       number = {1},
        pages = {388-411},
          doi = {10.1093/mnras/stt2204},
archivePrefix = {arXiv},
       eprint = {1311.2949},
 primaryClass = {astro-ph.CO},
       adsurl = {https://ui.adsabs.harvard.edu/abs/2014MNRAS.438..388M}
}

@ARTICLE{Schwartz04,
       author = {{Schwartz}, Colleen M. and {Martin}, Crystal L.},
        title = "{A Keck/HIRES Study of Kinematics of the Cold Interstellar Medium in Dwarf Starburst Galaxies}",
      journal = {\apj},
         year = 2004,
        month = jul,
       volume = {610},
       number = {1},
        pages = {201-212},
          doi = {10.1086/421546},
archivePrefix = {arXiv},
       eprint = {astro-ph/0404148},
 primaryClass = {astro-ph},
       adsurl = {https://ui.adsabs.harvard.edu/abs/2004ApJ...610..201S}
}

@ARTICLE{Grimes09,
       author = {{Grimes}, J.~P. and {Heckman}, T. and {Aloisi}, A. and {Calzetti}, D. and {Leitherer}, C. and {Martin}, C.~L. and {Meurer}, G. and {Sembach}, K. and {Strickland}, D.},
        title = "{Observations of Starburst Galaxies With Far-Ultraviolet Spectrographic Explorer: Galactic Feedback in the Local Universe}",
      journal = {\apjs},
         year = 2009,
        month = mar,
       volume = {181},
       number = {1},
        pages = {272-320},
          doi = {10.1088/0067-0049/181/1/272},
       adsurl = {https://ui.adsabs.harvard.edu/abs/2009ApJS..181..272G}
}

@ARTICLE{Chisholm15,
       author = {{Chisholm}, John and {Tremonti}, Christy A. and {Leitherer}, Claus and {Chen}, Yanmei and {Wofford}, Aida and {Lundgren}, Britt},
        title = "{Scaling Relations Between Warm Galactic Outflows and Their Host Galaxies}",
      journal = {\apj},
         year = 2015,
        month = oct,
       volume = {811},
       number = {2},
          eid = {149},
        pages = {149},
          doi = {10.1088/0004-637X/811/2/149},
archivePrefix = {arXiv},
       eprint = {1412.2139},
 primaryClass = {astro-ph.GA},
       adsurl = {https://ui.adsabs.harvard.edu/abs/2015ApJ...811..149C}
}

@ARTICLE{Karachentsev13,
       author = {{Karachentsev}, Igor D. and {Makarov}, Dmitry I. and {Kaisina}, Elena I.},
        title = "{Updated Nearby Galaxy Catalog}",
      journal = {\aj},
         year = 2013,
        month = apr,
       volume = {145},
       number = {4},
          eid = {101},
        pages = {101},
          doi = {10.1088/0004-6256/145/4/101},
archivePrefix = {arXiv},
       eprint = {1303.5328},
 primaryClass = {astro-ph.CO},
       adsurl = {https://ui.adsabs.harvard.edu/abs/2013AJ....145..101K}
}

@ARTICLE{Schmidt14,
       author = {{Schmidt}, P. and {J{\'o}zsa}, G.~I.~G. and {Gentile}, G. and {Oh}, S.-H. and {Schuberth}, Y. and {Ben Bekhti}, N. and {Winkel}, B. and {Klein}, U.},
        title = "{Structure and kinematics of the nearby dwarf galaxy UGCA 105}",
      journal = {\aap},
         year = 2014,
        month = jan,
       volume = {561},
          eid = {A28},
        pages = {A28},
          doi = {10.1051/0004-6361/201118170},
archivePrefix = {arXiv},
       eprint = {1311.4392},
 primaryClass = {astro-ph.GA},
       adsurl = {https://ui.adsabs.harvard.edu/abs/2014A&A...561A..28S}
}

@ARTICLE{Fox26,
       author = {{Fox}, Andrew J. and {Mishra}, Sapna and {Cashman}, Frances H. and {French}, David M. and {Richter}, Philipp and {Bordoloi}, Rongmon and {Lehner}, Nicolas and {Tumlinson}, Jason and {Borthakur}, Sanchayeeta},
        title = "{Low-metallicity Gas on the Outskirts of the Local Group: The Circumgalactic Medium of Sextans B}",
      journal = {\apj},
         year = 2026,
        month = mar,
       volume = {999},
       number = {1},
          eid = {107},
        pages = {107},
          doi = {10.3847/1538-4357/ae35fb},
archivePrefix = {arXiv},
       eprint = {2601.00437},
 primaryClass = {astro-ph.GA},
       adsurl = {https://ui.adsabs.harvard.edu/abs/2026ApJ...999..107F}
}

@ARTICLE{Zahedy21,
       author = {{Zahedy}, Fakhri S. and {Chen}, Hsiao-Wen and {Cooper}, Thomas M. and {Boettcher}, Erin and {Johnson}, Sean D. and {Rudie}, Gwen C. and {Chen}, Mandy C. and {Cantalupo}, Sebastiano and {Cooksey}, Kathy L. and {Faucher-Gigu{\`e}re}, Claude-Andr{\'e} and {Greene}, Jenny E. and {Lopez}, Sebastian and {Mulchaey}, John S. and {Penton}, Steven V. and {Petitjean}, Patrick and {Putman}, Mary E. and {Rafelski}, Marc and {Rauch}, Michael and {Schaye}, Joop and {Simcoe}, Robert A. and {Walth}, Gregory L.},
        title = "{The Cosmic Ultraviolet Baryon Survey (CUBS) - III. Physical properties and elemental abundances of Lyman-limit systems at z < 1}",
      journal = {\mnras},
         year = 2021,
        month = sep,
       volume = {506},
       number = {1},
        pages = {877-902},
          doi = {10.1093/mnras/stab1661},
archivePrefix = {arXiv},
       eprint = {2106.04608},
 primaryClass = {astro-ph.GA},
       adsurl = {https://ui.adsabs.harvard.edu/abs/2021MNRAS.506..877Z}
}

@ARTICLE{Johnson26,
       author = {{Johnson}, Sean D. and {Mishra}, Nishant and {Muzahid}, Sowgat and {Rudie}, Gwen C. and {Zahedy}, Fakhri S. and {Qu}, Zhijie and {Faucher-Gigu{\`e}re}, Claude-Andr{\'e} and {Stern}, Jonathan and {Li}, Jennifer I.-Hsiu and {Fuller}, Elise and {Cantalupo}, Sebastiano and {Chen}, Hsiao-Wen and {Kadri}, Ahmad and {Kumar}, Suyash and {Liu}, Zhuoqi (Will) and {Walth}, Gregory},
        title = "{MUSEQuBES: Physical Conditions, Origins, and Multielement Abundances of the Circumgalactic Medium of an Isolated, Star-forming Dwarf Galaxy at z = 0.57}",
      journal = {\apjl},
         year = 2026,
        month = jan,
       volume = {996},
       number = {2},
          eid = {L30},
        pages = {L30},
          doi = {10.3847/2041-8213/ae2f5f},
archivePrefix = {arXiv},
       eprint = {2510.06310},
 primaryClass = {astro-ph.GA},
       adsurl = {https://ui.adsabs.harvard.edu/abs/2026ApJ...996L..30J}
}

@ARTICLE{Mishra24,
       author = {{Mishra}, Nishant and {Johnson}, Sean D. and {Rudie}, Gwen C. and {Chen}, Hsiao-Wen and {Schaye}, Joop and {Qu}, Zhijie and {Zahedy}, Fakhri S. and {Boettcher}, Erin T. and {Cantalupo}, Sebastiano and {Chen}, Mandy C. and {Faucher-Gigu{\'e}re}, Claude-Andr{\'e} and {Greene}, Jenny E. and {Li}, Jennifer I.-Hsiu and {Liu}, Zhuoqi (Will) and {Lopez}, Sebastian and {Petitjean}, Patrick},
        title = "{The Cosmic Ultraviolet Baryon Survey (CUBS). IX. The Enriched Circumgalactic and Intergalactic Medium Around Star-forming Field Dwarf Galaxies Traced by O VI Absorption}",
      journal = {\apj},
         year = 2024,
        month = nov,
       volume = {976},
       number = {1},
          eid = {149},
        pages = {149},
          doi = {10.3847/1538-4357/ad7b0a},
archivePrefix = {arXiv},
       eprint = {2408.11151},
 primaryClass = {astro-ph.GA},
       adsurl = {https://ui.adsabs.harvard.edu/abs/2024ApJ...976..149M}
}

@ARTICLE{Johnson17,
       author = {{Johnson}, Sean D. and {Chen}, Hsiao-Wen and {Mulchaey}, John S. and {Schaye}, Joop and {Straka}, Lorrie A.},
        title = "{The Extent of Chemically Enriched Gas around Star-forming Dwarf Galaxies}",
      journal = {\apjl},
         year = 2017,
        month = nov,
       volume = {850},
       number = {1},
          eid = {L10},
        pages = {L10},
          doi = {10.3847/2041-8213/aa9370},
archivePrefix = {arXiv},
       eprint = {1710.06441},
 primaryClass = {astro-ph.GA},
       adsurl = {https://ui.adsabs.harvard.edu/abs/2017ApJ...850L..10J}
}

@ARTICLE{Fox19,
       author = {{Fox}, Andrew J. and {Richter}, Philipp and {Ashley}, Trisha and {Heckman}, Timothy M. and {Lehner}, Nicolas and {Werk}, Jessica K. and {Bordoloi}, Rongmon and {Peeples}, Molly S.},
        title = "{The Mass Inflow and Outflow Rates of the Milky Way}",
      journal = {\apj},
         year = 2019,
        month = oct,
       volume = {884},
       number = {1},
          eid = {53},
        pages = {53},
          doi = {10.3847/1538-4357/ab40ad},
archivePrefix = {arXiv},
       eprint = {1909.05561},
 primaryClass = {astro-ph.GA},
       adsurl = {https://ui.adsabs.harvard.edu/abs/2019ApJ...884...53F}
}

@ARTICLE{Kobulnicky96,
       author = {{Kobulnicky}, Henry A. and {Skillman}, Evan D.},
        title = "{Elemental Abundance Variations and Chemical Enrichment from Massive Stars in Starbursts. I. NGC 4214}",
      journal = {\apj},
         year = 1996,
        month = nov,
       volume = {471},
        pages = {211},
          doi = {10.1086/177964},
       adsurl = {https://ui.adsabs.harvard.edu/abs/1996ApJ...471..211K}
}

@ARTICLE{Asplund09,
       author = {{Asplund}, Martin and {Grevesse}, Nicolas and {Sauval}, A. Jacques and {Scott}, Pat},
        title = "{The Chemical Composition of the Sun}",
      journal = {\araa},
         year = 2009,
        month = sep,
       volume = {47},
       number = {1},
        pages = {481-522},
          doi = {10.1146/annurev.astro.46.060407.145222},
archivePrefix = {arXiv},
       eprint = {0909.0948},
 primaryClass = {astro-ph.SR},
       adsurl = {https://ui.adsabs.harvard.edu/abs/2009ARA&A..47..481A}
}

@ARTICLE{Fox16,
       author = {{Fox}, Andrew J. and {Lehner}, Nicolas and {Lockman}, Felix J. and {Wakker}, Bart P. and {Hill}, Alex S. and {Heitsch}, Fabian and {Stark}, David V. and {Barger}, Kathleen A. and {Sembach}, Kenneth R. and {Rahman}, Mubdi},
        title = "{On the Metallicity and Origin of the Smith High-velocity Cloud}",
      journal = {\apjl},
         year = 2016,
        month = jan,
       volume = {816},
       number = {1},
          eid = {L11},
        pages = {L11},
          doi = {10.3847/2041-8205/816/1/L11},
archivePrefix = {arXiv},
       eprint = {1512.04957},
 primaryClass = {astro-ph.GA},
       adsurl = {https://ui.adsabs.harvard.edu/abs/2016ApJ...816L..11F}
}

@ARTICLE{HI4PI16,
       author = {{HI4PI Collaboration} and {Ben Bekhti}, N. and {Fl{\"o}er}, L. and {Keller}, R. and {Kerp}, J. and {Lenz}, D. and {Winkel}, B. and {Bailin}, J. and {Calabretta}, M.~R. and {Dedes}, L. and {Ford}, H.~A. and {Gibson}, B.~K. and {Haud}, U. and {Janowiecki}, S. and {Kalberla}, P.~M.~W. and {Lockman}, F.~J. and {McClure-Griffiths}, N.~M. and {Murphy}, T. and {Nakanishi}, H. and {Pisano}, D.~J. and {Staveley-Smith}, L.},
        title = "{HI4PI: A full-sky H I survey based on EBHIS and GASS}",
      journal = {\aap},
         year = 2016,
        month = oct,
       volume = {594},
          eid = {A116},
        pages = {A116},
          doi = {10.1051/0004-6361/201629178},
archivePrefix = {arXiv},
       eprint = {1610.06175},
 primaryClass = {astro-ph.GA},
       adsurl = {https://ui.adsabs.harvard.edu/abs/2016A&A...594A.116H}
}

@ARTICLE{Wakker15,
       author = {{Wakker}, Bart P. and {Hernandez}, Audra K. and {French}, David M. and {Kim}, Tae-Sun and {Oppenheimer}, Benjamin D. and {Savage}, Blair D.},
        title = "{Nearby Galaxy Filaments and the Ly-alpha Forest: Confronting Simulations and the UV Background with Observations}",
      journal = {\apj},
         year = 2015,
        month = nov,
       volume = {814},
       number = {1},
          eid = {40},
        pages = {40},
          doi = {10.1088/0004-637X/814/1/40},
archivePrefix = {arXiv},
       eprint = {1504.02539},
 primaryClass = {astro-ph.CO},
       adsurl = {https://ui.adsabs.harvard.edu/abs/2015ApJ...814...40W}
}

@ARTICLE{Jiang26,
       author = {{Jiang}, Xiaochuan and {Fang}, Taotao and {Yan}, Shulan},
        title = "{The Vertical Structure and Asymmetry of Mg ii-enriched Gas in the Milky Way Disk}",
      journal = {Science China Physics, Mechanics, and Astronomy},
         year = 2026,
        month = jul,
        volume = {69},
        number = {7},
        eid = {279512},
        pages = {279512},
        doi = {10.1007/s11433-025-2965-5},
        archivePrefix = {arXiv},
archivePrefix = {arXiv},
       eprint = {2604.23686},
 primaryClass = {astro-ph.GA},
       adsurl = {https://ui.adsabs.harvard.edu/abs/2026arXiv260423686J}
}

@ARTICLE{Davies20,
       author = {{Davies}, Jonathan J. and {Crain}, Robert A. and {Oppenheimer}, Benjamin D. and {Schaye}, Joop},
        title = "{The quenching and morphological evolution of central galaxies is facilitated by the feedback-driven expulsion of circumgalactic gas}",
      journal = {\mnras},
         year = 2020,
        month = jan,
       volume = {491},
       number = {3},
        pages = {4462-4480},
          doi = {10.1093/mnras/stz3201},
archivePrefix = {arXiv},
       eprint = {1908.11380},
 primaryClass = {astro-ph.GA},
       adsurl = {https://ui.adsabs.harvard.edu/abs/2020MNRAS.491.4462D}
}

@ARTICLE{DESI2026,
       author = {{DESI Collaboration} and {Abdul Karim}, M. and {Adame}, A.~G. and {Aguado}, D. and {Aguilar}, J. and {Ahlen}, S. and {Alam}, S. and {Aldering}, G. and {Alexander}, D.~M. and {Alfarsy}, R. and {Allen}, L. and {Allende Prieto}, C. and {Alves}, O. and {Anand}, A. and {Andrade}, U. and {Armengaud}, E. and {Avila}, S. and {Aviles}, A. and {Awan}, H. and {Bailey}, S. and {Baleato Lizancos}, A. and {Ballester}, O. and {Bault}, A. and {Bautista}, J. and {Bean}, R. and {Behera}, J. and {BenZvi}, S. and {Beraldo e Silva}, L. and {Bermejo-Climent}, J.~R. and {Beutler}, F. and {Bianchi}, D. and {Blake}, C. and {Blum}, R. and {Bolton}, A.~S. and {Bonici}, M. and {Brieden}, S. and {Brodzeller}, A. and {Brooks}, D. and {Buckley-Geer}, E. and {Burtin}, E. and {Bystr{\"o}m}, A. and {Canning}, R. and {Carnero Rosell}, A. and {Carr}, A. and {Carrilho}, P. and {Casas}, L. and {Castander}, F.~J. and {Cereskaite}, R. and {Cervantes-Cota}, J.~L. and {Chaussidon}, E. and {Chaves-Montero}, J. and {Chen}, S. and {Chen}, X. and {Circosta}, C. and {Claybaugh}, T. and {Cole}, S. and {Cooper}, A.~P. and {Cousinou}, M.-C. and {Cuceu}, A. and {Davis}, T.~M. and {Dawson}, K.~S. and {de Belsunce}, R. and {de la Cruz}, R. and {de la Macorra}, A. and {de Mattia}, A. and {Deiosso}, N. and {Della Costa}, J. and {Demina}, R. and {Demirbozan}, U. and {DeRose}, J. and {Dey}, A. and {Dey}, B. and {Ding}, J. and {Ding}, Z. and {Doel}, P. and {Douglass}, K. and {Dowicz}, M. and {Ebina}, H. and {Edelstein}, J. and {Eisenstein}, D.~J. and {Elbers}, W. and {Emas}, N. and {Escoffier}, S. and {Fagrelius}, P. and {Fan}, X. and {Fanning}, K. and {Favole}, G. and {Fawcett}, V.~A. and {Fern{\'a}ndez-Garc{\'\i}a}, E. and {Ferraro}, S. and {Findlay}, N. and {Font-Ribera}, A. and {Forero-Romero}, J.~E. and {Forero-S{\'a}nchez}, D. and {Frenk}, C.~S. and {G{\"a}nsicke}, B.~T. and {Galbany}, L. and {Garc{\'\i}a-Bellido}, J. and {Garcia-Quintero}, C. and {Garrison}, L.~H. and {Gazta{\~n}aga}, E. and {Gil-Mar{\'\i}n}, H. and {Gloudemans}, A. and {Gnedin}, O.~Y. and {Gontcho A Gontcho}, S. and {Gonzalez}, D. and {Gonzalez-Morales}, A.~X. and {Gonzalez-Perez}, V. and {Gordon}, C. and {Graur}, O. and {Green}, D. and {Gruen}, D. and {Gsponer}, R. and {Guandalin}, C. and {Gutierrez}, G. and {Guy}, J. and {Hahn}, C. and {Han}, J.~J. and {Han}, J. and {He}, S. and {Herrera-Alcantar}, H.~K. and {Heydenreich}, S. and {Honscheid}, K. and {Hou}, J. and {Howlett}, C. and {Huterer}, D. and {Ir{\v{s}}i{\v{c}}}, V. and {Ishak}, M. and {Jacques}, A. and {Jiang}, L. and {Jimenez}, J. and {Jing}, Y.~P. and {Joachimi}, B. and {Joudaki}, S. and {Joyce}, R. and {Jullo}, E. and {Juneau}, S. and {Kara{\c{c}}ayl{\i}}, N.~G. and {Karim}, T. and {Kehoe}, R. and {Kent}, S. and {Khederlarian}, A. and {Kirkby}, D. and {Kisner}, T. and {Kitaura}, F.-S. and {Kizhuprakkat}, N. and {Kong}, H. and {Koposov}, S.~E. and {Kremin}, A. and {Krolewski}, A. and {Lahav}, O. and {Lai}, Y. and {Lamman}, C. and {Lan}, T.-W. and {Landriau}, M. and {Lang}, D. and {Lange}, J.~U. and {Lasker}, J. and {Le Goff}, J.~M. and {Le Guillou}, L. and {Leauthaud}, A. and {Levi}, M.~E. and {Li}, S. and {Li}, T.~S. and {Liu}, W. and {Lodha}, K. and {Lokken}, M. and {Luo}, Y. and {Magneville}, C. and {Manera}, M. and {Manser}, C.~J. and {Margala}, D. and {Martini}, P. and {Maus}, M. and {McCullough}, J. and {McDonald}, P. and {Medina}, G.~E. and {Medina-Varela}, L. and {Meisner}, A. and {Mena-Fern{\'a}ndez}, J. and {Menegas}, A. and {Meneses-Rizo}, J. and {Mezcua}, M. and {Miquel}, R. and {Montero-Camacho}, P. and {Moon}, J. and {Moustakas}, J. and {Mu{\~n}oz-Guti{\'e}rrez}, A. and {Mu noz-Santos}, D. and {Myers}, A.~D. and {Myles}, J. and {Nadathur}, S. and {Najita}, J. and {Napolitano}, L. and {Newman}, J.~A. and {Nikakhtar}, F. and {Nikutta}, R. and {Niz}, G. and {Noriega}, H.~E. and {Nugent}, P.},
        title = "{Data Release 1 of the Dark Energy Spectroscopic Instrument}",
      journal = {\aj},
         year = 2026,
        month = may,
       volume = {171},
       number = {5},
          eid = {285},
        pages = {285},
          doi = {10.3847/1538-3881/ae4c43},
archivePrefix = {arXiv},
       eprint = {2503.14745},
 primaryClass = {astro-ph.CO},
       adsurl = {https://ui.adsabs.harvard.edu/abs/2026AJ....171..285D}
}

@ARTICLE{Meyer17,
       author = {{Meyer}, Martin and {Robotham}, Aaron and {Obreschkow}, Danail and {Westmeier}, Tobias and {Duffy}, Alan R. and {Staveley-Smith}, Lister},
        title = "{Tracing HI Beyond the Local Universe}",
      journal = {\pasa},
         year = 2017,
        month = nov,
       volume = {34},
        pages = {52},
          doi = {10.1017/pasa.2017.31},
archivePrefix = {arXiv},
       eprint = {1705.04210},
 primaryClass = {astro-ph.CO},
       adsurl = {https://ui.adsabs.harvard.edu/abs/2017PASA...34...52M}
}

@ARTICLE{Fouque90,
       author = {{Fouque}, P. and {Durand}, N. and {Bottinelli}, L. and {Gouguenheim}, L. and {Paturel}, G.},
        title = "{An HI survey of late-type galaxies in the southern hemisphere. I. The SGC sample.}",
      journal = {\aaps},
         year = 1990,
        month = dec,
       volume = {86},
        pages = {473},
       adsurl = {https://ui.adsabs.harvard.edu/abs/1990A&AS...86..473F}
}

@ARTICLE{Koribalski04,
       author = {{Koribalski}, B.~S. and {Staveley-Smith}, L. and {Kilborn}, V.~A. and {Ryder}, S.~D. and {Kraan-Korteweg}, R.~C. and {Ryan-Weber}, E.~V. and {Ekers}, R.~D. and {Jerjen}, H. and {Henning}, P.~A. and {Putman}, M.~E. and {Zwaan}, M.~A. and {de Blok}, W.~J.~G. and {Calabretta}, M.~R. and {Disney}, M.~J. and {Minchin}, R.~F. and {Bhathal}, R. and {Boyce}, P.~J. and {Drinkwater}, M.~J. and {Freeman}, K.~C. and {Gibson}, B.~K. and {Green}, A.~J. and {Haynes}, R.~F. and {Juraszek}, S. and {Kesteven}, M.~J. and {Knezek}, P.~M. and {Mader}, S. and {Marquarding}, M. and {Meyer}, M. and {Mould}, J.~R. and {Oosterloo}, T. and {O'Brien}, J. and {Price}, R.~M. and {Sadler}, E.~M. and {Schr{\"o}der}, A. and {Stewart}, I.~M. and {Stootman}, F. and {Waugh}, M. and {Warren}, B.~E. and {Webster}, R.~L. and {Wright}, A.~E.},
        title = "{The 1000 Brightest HIPASS Galaxies: H I Properties}",
      journal = {\aj},
         year = 2004,
        month = jul,
       volume = {128},
       number = {1},
        pages = {16-46},
          doi = {10.1086/421744},
archivePrefix = {arXiv},
       eprint = {astro-ph/0404436},
 primaryClass = {astro-ph},
       adsurl = {https://ui.adsabs.harvard.edu/abs/2004AJ....128...16K}
}

@ARTICLE{Westmeier21,
       author = {{Westmeier}, T. and {Kitaeff}, S. and {Pallot}, D. and {Serra}, P. and {van der Hulst}, J.~M. and {Jurek}, R.~J. and {Elagali}, A. and {For}, B.-Q. and {Kleiner}, D. and {Koribalski}, B.~S. and {Lee-Waddell}, K. and {Mould}, J.~R. and {Reynolds}, T.~N. and {Rhee}, J. and {Staveley-Smith}, L.},
        title = "{SOFIA 2 - an automated, parallel H I source finding pipeline for the WALLABY survey}",
      journal = {\mnras},
         year = 2021,
        month = sep,
       volume = {506},
       number = {3},
        pages = {3962-3976},
          doi = {10.1093/mnras/stab1881},
archivePrefix = {arXiv},
       eprint = {2106.15789},
 primaryClass = {astro-ph.IM},
       adsurl = {https://ui.adsabs.harvard.edu/abs/2021MNRAS.506.3962W}
}

@ARTICLE{Westmeier22,
       author = {{Westmeier}, T. and {Deg}, N. and {Spekkens}, K. and {Reynolds}, T.~N. and {Shen}, A.~X. and {Gaudet}, S. and {Goliath}, S. and {Huynh}, M.~T. and {Venkataraman}, P. and {Lin}, X. and {O'Beirne}, T. and {Catinella}, B. and {Cortese}, L. and {D{\'e}nes}, H. and {Elagali}, A. and {For}, B.-Q. and {J{\'o}zsa}, G.~I.~G. and {Howlett}, C. and {van der Hulst}, J.~M. and {Jurek}, R.~J. and {Kamphuis}, P. and {Kilborn}, V.~A. and {Kleiner}, D. and {Koribalski}, B.~S. and {Lee-Waddell}, K. and {Murugeshan}, C. and {Rhee}, J. and {Serra}, P. and {Shao}, L. and {Staveley-Smith}, L. and {Wang}, J. and {Wong}, O.~I. and {Zwaan}, M.~A. and {Allison}, J.~R. and {Anderson}, C.~S. and {Ball}, Lewis and {Bock}, D.~C.-J. and {Brodrick}, D. and {Bunton}, J.~D. and {Cooray}, F.~R. and {Gupta}, N. and {Hayman}, D.~B. and {Mahony}, E.~K. and {Moss}, V.~A. and {Ng}, A. and {Pearce}, S.~E. and {Raja}, W. and {Roxby}, D.~N. and {Voronkov}, M.~A. and {Warhurst}, K.~A. and {Courtois}, H.~M. and {Said}, K.},
        title = "{WALLABY pilot survey: Public release of H I data for almost 600 galaxies from phase 1 of ASKAP pilot observations}",
      journal = {\pasa},
         year = 2022,
        month = nov,
       volume = {39},
          eid = {e058},
        pages = {e058},
          doi = {10.1017/pasa.2022.50},
archivePrefix = {arXiv},
       eprint = {2211.07094},
 primaryClass = {astro-ph.GA},
       adsurl = {https://ui.adsabs.harvard.edu/abs/2022PASA...39...58W}
}

@ARTICLE{Serra15,
       author = {{Serra}, Paolo and {Westmeier}, Tobias and {Giese}, Nadine and {Jurek}, Russell and {Fl{\"o}er}, Lars and {Popping}, Attila and {Winkel}, Benjamin and {van der Hulst}, Thijs and {Meyer}, Martin and {Koribalski}, B{\"a}rbel S. and {Staveley-Smith}, Lister and {Courtois}, H{\'e}l{\`e}ne},
        title = "{SOFIA: a flexible source finder for 3D spectral line data}",
      journal = {\mnras},
         year = 2015,
        month = apr,
       volume = {448},
       number = {2},
        pages = {1922-1929},
          doi = {10.1093/mnras/stv079},
archivePrefix = {arXiv},
       eprint = {1501.03906},
 primaryClass = {astro-ph.IM},
       adsurl = {https://ui.adsabs.harvard.edu/abs/2015MNRAS.448.1922S}
}

@ARTICLE{Walter08,
       author = {{Walter}, Fabian and {Brinks}, Elias and {de Blok}, W.~J.~G. and {Bigiel}, Frank and {Kennicutt}, Jr., Robert C. and {Thornley}, Michele D. and {Leroy}, Adam},
        title = "{THINGS: The H I Nearby Galaxy Survey}",
      journal = {\aj},
         year = 2008,
        month = dec,
       volume = {136},
       number = {6},
        pages = {2563-2647},
          doi = {10.1088/0004-6256/136/6/2563},
archivePrefix = {arXiv},
       eprint = {0810.2125},
 primaryClass = {astro-ph},
       adsurl = {https://ui.adsabs.harvard.edu/abs/2008AJ....136.2563W}
}

@ARTICLE{Sylos25,
       author = {{Sylos Labini}, Francesco and {De Marzo}, Giordano and {Straccamore}, Matteo},
        title = "{Breaking the Degeneracy between Warps and Radial Flows in External Galaxies}",
      journal = {\apj},
         year = 2025,
        month = jul,
       volume = {988},
       number = {1},
          eid = {122},
        pages = {122},
          doi = {10.3847/1538-4357/adc71c},
archivePrefix = {arXiv},
       eprint = {2503.22306},
 primaryClass = {astro-ph.GA},
       adsurl = {https://ui.adsabs.harvard.edu/abs/2025ApJ...988..122S}
}

@ARTICLE{Binney92,
       author = {{Binney}, James},
        title = "{Warps.}",
      journal = {\araa},
         year = 1992,
        month = jan,
       volume = {30},
        pages = {51-74},
          doi = {10.1146/annurev.aa.30.090192.000411},
       adsurl = {https://ui.adsabs.harvard.edu/abs/1992ARA&A..30...51B}
}

@ARTICLE{Visser80,
       author = {{Visser}, H.~C.~D.},
        title = "{The dynamics of the spiral galaxy M 81. II. Gas dynamics and neutral-hydrogen observations.}",
      journal = {\aap},
         year = 1980,
        month = aug,
       volume = {88},
        pages = {159-174},
       adsurl = {https://ui.adsabs.harvard.edu/abs/1980A&A....88..159V}
}

@ARTICLE{Cao22,
       author = {{Cao}, Xiao and {Chen}, Yan-Mei and {Shi}, Yong and {Bao}, Min and {Moiseev}, Alexei and {Bizyaev}, Dmitry and {Li}, Song-Lin and {Fern{\'a}ndez-Trincado}, Jos{\'e} G. and {Riffel}, Rogemar A. and {Riffel}, Rog{\'e}rio and {Lane}, Richard R.},
        title = "{Multiple gas acquisition events in galaxies with dual misaligned gas disks}",
      journal = {Nature Astronomy},
         year = 2022,
        month = dec,
       volume = {6},
        pages = {1464-1472},
          doi = {10.1038/s41550-022-01788-8},
archivePrefix = {arXiv},
       eprint = {2209.15417},
 primaryClass = {astro-ph.GA},
       adsurl = {https://ui.adsabs.harvard.edu/abs/2022NatAs...6.1464C}
}

@INPROCEEDINGS{Fraternali17,
       author = {{Fraternali}, Filippo},
        title = "{Gas Accretion via Condensation and Fountains}",
    booktitle = {Gas Accretion onto Galaxies},
         year = 2017,
       editor = {{Fox}, Andrew and {Dav{\'e}}, Romeel},
       series = {Astrophysics and Space Science Library},
       volume = {430},
        month = jan,
        pages = {323},
          doi = {10.1007/978-3-319-52512-9_14},
archivePrefix = {arXiv},
       eprint = {1612.00477},
 primaryClass = {astro-ph.GA},
       adsurl = {https://ui.adsabs.harvard.edu/abs/2017ASSL..430..323F}
}

@ARTICLE{Marinacci10,
       author = {{Marinacci}, Federico and {Binney}, James and {Fraternali}, Filippo and {Nipoti}, Carlo and {Ciotti}, Luca and {Londrillo}, Pasquale},
        title = "{The mode of gas accretion on to star-forming galaxies}",
      journal = {\mnras},
         year = 2010,
        month = may,
       volume = {404},
       number = {3},
        pages = {1464-1474},
          doi = {10.1111/j.1365-2966.2010.16352.x},
archivePrefix = {arXiv},
       eprint = {1001.2446},
 primaryClass = {astro-ph.GA},
       adsurl = {https://ui.adsabs.harvard.edu/abs/2010MNRAS.404.1464M}
}

@ARTICLE{Fraternali08,
       author = {{Fraternali}, F. and {Binney}, J.~J.},
        title = "{Accretion of gas on to nearby spiral galaxies}",
      journal = {\mnras},
         year = 2008,
        month = may,
       volume = {386},
       number = {2},
        pages = {935-944},
          doi = {10.1111/j.1365-2966.2008.13071.x},
archivePrefix = {arXiv},
       eprint = {0802.0496},
 primaryClass = {astro-ph},
       adsurl = {https://ui.adsabs.harvard.edu/abs/2008MNRAS.386..935F}
}

@ARTICLE{Iwamoto99,
       author = {{Iwamoto}, Koichi and {Brachwitz}, Franziska and {Nomoto}, Ken'ICHI and {Kishimoto}, Nobuhiro and {Umeda}, Hideyuki and {Hix}, W. Raphael and {Thielemann}, Friedrich-Karl},
        title = "{Nucleosynthesis in Chandrasekhar Mass Models for Type IA Supernovae and Constraints on Progenitor Systems and Burning-Front Propagation}",
      journal = {\apjs},
         year = 1999,
        month = dec,
       volume = {125},
       number = {2},
        pages = {439-462},
          doi = {10.1086/313278},
archivePrefix = {arXiv},
       eprint = {astro-ph/0002337},
 primaryClass = {astro-ph},
       adsurl = {https://ui.adsabs.harvard.edu/abs/1999ApJS..125..439I}
}

@ARTICLE{Johnson19,
       author = {{Johnson}, Jennifer A.},
        title = "{Populating the periodic table: Nucleosynthesis of the elements}",
      journal = {Science},
         year = 2019,
        month = feb,
       volume = {363},
       number = {6426},
        pages = {474-478},
          doi = {10.1126/science.aau9540},
       adsurl = {https://ui.adsabs.harvard.edu/abs/2019Sci...363..474J}
}

@ARTICLE{Srivastava11,
       author = {{Srivastava}, Shweta and {Kantharia}, N.~G. and {Srivastava}, D.~C.},
        title = "{GMRT Low Radio Frequency Study of the Wolf Rayet Galaxy NGC 4214 and Detection of a Distant Galaxy}",
      journal = {Journal of Astrophysics and Astronomy},
         year = 2011,
        month = dec,
       volume = {32},
       number = {4},
        pages = {465-466},
          doi = {10.1007/s12036-011-9093-5},
       adsurl = {https://ui.adsabs.harvard.edu/abs/2011JApA...32..465S}
}

@ARTICLE{MasHesse99,
       author = {{Mas-Hesse}, J. Miguel and {Kunth}, Daniel},
        title = "{A comprehensive study of intense star formation bursts in irregular and compact galaxies}",
      journal = {\aap},
         year = 1999,
        month = sep,
       volume = {349},
        pages = {765-795},
          doi = {10.48550/arXiv.astro-ph/9812072},
archivePrefix = {arXiv},
       eprint = {astro-ph/9812072},
 primaryClass = {astro-ph},
       adsurl = {https://ui.adsabs.harvard.edu/abs/1999A&A...349..765M}
}

@ARTICLE{Sargent91,
       author = {{Sargent}, Wallace L.~W. and {Filippenko}, Alexei V.},
        title = "{Luminous Clusters of Wolf-Rayet Stars in the SBmIII Galaxy NGC 4214}",
      journal = {\aj},
         year = 1991,
        month = jul,
       volume = {102},
        pages = {107},
          doi = {10.1086/115859},
       adsurl = {https://ui.adsabs.harvard.edu/abs/1991AJ....102..107S}
}

@ARTICLE{Leitherer96,
       author = {{Leitherer}, Claus and {Vacca}, William D. and {Conti}, Peter S. and {Filippenko}, Alexei V. and {Robert}, Carmelle and {Sargent}, Wallace L.~W.},
        title = "{Hubble Space Telescope Ultraviolet Imaging and Spectroscopy of the Bright Starburst in the Wolf-Rayet Galaxy NGC 4214}",
      journal = {\apj},
         year = 1996,
        month = jul,
       volume = {465},
        pages = {717},
          doi = {10.1086/177456},
       adsurl = {https://ui.adsabs.harvard.edu/abs/1996ApJ...465..717L}
}

@ARTICLE{Drozdovsky02,
       author = {{Drozdovsky}, Igor O. and {Schulte-Ladbeck}, Regina E. and {Hopp}, Ulrich and {Greggio}, Laura and {Crone}, Mary M.},
        title = "{The Dwarf Irregular/Wolf-Rayet Galaxy NGC 4214. I. A New Distance, Stellar Content, and Global Parameters}",
      journal = {\aj},
         year = 2002,
        month = aug,
       volume = {124},
       number = {2},
        pages = {811-827},
          doi = {10.1086/341165},
archivePrefix = {arXiv},
       eprint = {astro-ph/0205011},
 primaryClass = {astro-ph},
       adsurl = {https://ui.adsabs.harvard.edu/abs/2002AJ....124..811D}
}

@ARTICLE{Xu25,
       author = {{Xu}, Jin-Long and {Zhu}, Ming and {Yu}, Nai-Ping and {Zhang}, Chuan-Peng and {Liu}, Xiao-Lan and {Ai}, Mei and {Jiang}, Peng},
        title = "{FAST Discovery of a Gas-rich and Ultrafaint Dwarf Galaxy: KK153}",
      journal = {\apjl},
         year = 2025,
        month = apr,
       volume = {982},
       number = {2},
          eid = {L36},
        pages = {L36},
          doi = {10.3847/2041-8213/adbe7e},
archivePrefix = {arXiv},
       eprint = {2503.08999},
 primaryClass = {astro-ph.GA},
       adsurl = {https://ui.adsabs.harvard.edu/abs/2025ApJ...982L..36X}
}

@ARTICLE{Werk14,
       author = {{Werk}, Jessica K. and {Prochaska}, J. Xavier and {Tumlinson}, Jason and {Peeples}, Molly S. and {Tripp}, Todd M. and {Fox}, Andrew J. and {Lehner}, Nicolas and {Thom}, Christopher and {O'Meara}, John M. and {Ford}, Amanda Brady and {Bordoloi}, Rongmon and {Katz}, Neal and {Tejos}, Nicolas and {Oppenheimer}, Benjamin D. and {Dav{\'e}}, Romeel and {Weinberg}, David H.},
        title = "{The COS-Halos Survey: Physical Conditions and Baryonic Mass in the Low-redshift Circumgalactic Medium}",
      journal = {\apj},
         year = 2014,
        month = sep,
       volume = {792},
       number = {1},
          eid = {8},
        pages = {8},
          doi = {10.1088/0004-637X/792/1/8},
archivePrefix = {arXiv},
       eprint = {1403.0947},
 primaryClass = {astro-ph.CO},
       adsurl = {https://ui.adsabs.harvard.edu/abs/2014ApJ...792....8W}
}

@ARTICLE{Lehner13,
       author = {{Lehner}, N. and {Howk}, J.~C. and {Tripp}, T.~M. and {Tumlinson}, J. and {Prochaska}, J.~X. and {O'Meara}, J.~M. and {Thom}, C. and {Werk}, J.~K. and {Fox}, A.~J. and {Ribaudo}, J.},
        title = "{The Bimodal Metallicity Distribution of the Cool Circumgalactic Medium at z <\raisebox{-0.5ex}\textasciitilde 1}",
      journal = {\apj},
         year = 2013,
        month = jun,
       volume = {770},
       number = {2},
          eid = {138},
        pages = {138},
          doi = {10.1088/0004-637X/770/2/138},
archivePrefix = {arXiv},
       eprint = {1302.5424},
 primaryClass = {astro-ph.CO},
       adsurl = {https://ui.adsabs.harvard.edu/abs/2013ApJ...770..138L}
}

@ARTICLE{Gnat07,
       author = {{Gnat}, Orly and {Sternberg}, Amiel},
        title = "{Time-dependent Ionization in Radiatively Cooling Gas}",
      journal = {\apjs},
         year = 2007,
        month = feb,
       volume = {168},
       number = {2},
        pages = {213-230},
          doi = {10.1086/509786},
archivePrefix = {arXiv},
       eprint = {astro-ph/0608181},
 primaryClass = {astro-ph},
       adsurl = {https://ui.adsabs.harvard.edu/abs/2007ApJS..168..213G}
}

@ARTICLE{Tumlinson17,
       author = {{Tumlinson}, Jason and {Peeples}, Molly S. and {Werk}, Jessica K.},
        title = "{The Circumgalactic Medium}",
      journal = {\araa},
         year = 2017,
        month = aug,
       volume = {55},
       number = {1},
        pages = {389-432},
          doi = {10.1146/annurev-astro-091916-055240},
archivePrefix = {arXiv},
       eprint = {1709.09180},
 primaryClass = {astro-ph.GA},
       adsurl = {https://ui.adsabs.harvard.edu/abs/2017ARA&A..55..389T}
}

@ARTICLE{James14,
       author = {{James}, B.~L. and {Aloisi}, A. and {Heckman}, T. and {Sohn}, S.~T. and {Wolfe}, M.~A.},
        title = "{Investigating Nearby Star-forming Galaxies in the Ultraviolet with HST/COS Spectroscopy. I. Spectral Analysis and Interstellar Abundance Determinations}",
      journal = {\apj},
         year = 2014,
        month = nov,
       volume = {795},
       number = {2},
          eid = {109},
        pages = {109},
          doi = {10.1088/0004-637X/795/2/109},
archivePrefix = {arXiv},
       eprint = {1408.4420},
 primaryClass = {astro-ph.GA},
       adsurl = {https://ui.adsabs.harvard.edu/abs/2014ApJ...795..109J}
}

@ARTICLE{Jiang20,
       author = {{Jiang}, Peng and {Tang}, Ning-Yu and {Hou}, Li-Gang and {Liu}, Meng-Ting and {Kr{\v{c}}o}, Marko and {Qian}, Lei and {Sun}, Jing-Hai and {Ching}, Tao-Chung and {Liu}, Bin and {Duan}, Yan and {Yue}, You-Ling and {Gan}, Heng-Qian and {Yao}, Rui and {Li}, Hui and {Pan}, Gao-Feng and {Yu}, Dong-Jun and {Liu}, Hong-Fei and {Li}, Di and {Peng}, Bo and {Yan}, Jun and {FAST Collaboration}},
        title = "{The fundamental performance of FAST with 19-beam receiver at L band}",
      journal = {Research in Astronomy and Astrophysics},
         year = 2020,
        month = may,
       volume = {20},
       number = {5},
          eid = {064},
        pages = {064},
          doi = {10.1088/1674-4527/20/5/64},
archivePrefix = {arXiv},
       eprint = {2002.01786},
 primaryClass = {astro-ph.IM},
       adsurl = {https://ui.adsabs.harvard.edu/abs/2020RAA....20...64J}
}

@ARTICLE{Jiang19,
       author = {{Jiang}, Peng and {Yue}, YouLing and {Gan}, HengQian and {Yao}, Rui and {Li}, Hui and {Pan}, GaoFeng and {Sun}, JingHai and {Yu}, DongJun and {Liu}, HongFei and {Tang}, NingYu and {Qian}, Lei and {Lu}, JiGuang and {Yan}, Jun and {Peng}, Bo and {Zhang}, ShuXin and {Wang}, QiMing and {Li}, Qi and {Li}, Di and {FAST Collaboration}},
        title = "{Commissioning progress of the FAST}",
      journal = {Science China Physics, Mechanics, and Astronomy},
         year = 2019,
        month = may,
       volume = {62},
       number = {5},
          eid = {959502},
        pages = {959502},
          doi = {10.1007/s11433-018-9376-1},
archivePrefix = {arXiv},
       eprint = {1903.06324},
 primaryClass = {astro-ph.IM},
       adsurl = {https://ui.adsabs.harvard.edu/abs/2019SCPMA..6259502J}
}

@ARTICLE{Lin25,
       author = {{Lin}, Zeren and {Martin}, D. Christopher and {Matuszewski}, Mateusz and {Neill}, James D. and {Miles}, Drew M. and {Picouet}, Vincent and {Prusinski}, Nikolaus Z. and {Hoadley}, Keri and {McGurk}, Rosalie},
        title = "{Kinematically Complex Circumgalactic Gas Around a Low-mass Galaxy: Filamentary Inflow and Counterrotation in J0910b}",
      journal = {\apj},
         year = 2025,
        month = dec,
       volume = {995},
       number = {1},
          eid = {12},
        pages = {12},
          doi = {10.3847/1538-4357/ae10b2},
       adsurl = {https://ui.adsabs.harvard.edu/abs/2025ApJ...995...12L}
}

@ARTICLE{Warren12,
       author = {{Warren}, Steven R. and {Skillman}, Evan D. and {Stilp}, Adrienne M. and {Dalcanton}, Julianne J. and {Ott}, J{\"u}rgen and {Walter}, Fabian and {Petersen}, Eric A. and {Koribalski}, B{\"a}rbel and {West}, Andrew A.},
        title = "{Tracing Cold H I Gas in nearby, Low-mass Galaxies}",
      journal = {\apj},
         year = 2012,
        month = sep,
       volume = {757},
       number = {1},
          eid = {84},
        pages = {84},
          doi = {10.1088/0004-637X/757/1/84},
archivePrefix = {arXiv},
       eprint = {1208.0983},
 primaryClass = {astro-ph.CO},
       adsurl = {https://ui.adsabs.harvard.edu/abs/2012ApJ...757...84W}
}

@ARTICLE{Sanchez23,
       author = {{S{\'a}nchez-Cruces}, M. and {Rosado}, M.},
        title = "{H{\ensuremath{\alpha}} kinematics of superbubbles and supernova remnants of the dwarf galaxy NGC 4214}",
      journal = {\mnras},
         year = 2023,
        month = oct,
       volume = {524},
       number = {4},
        pages = {4907-4922},
          doi = {10.1093/mnras/stad2051},
archivePrefix = {arXiv},
       eprint = {2307.03207},
 primaryClass = {astro-ph.GA},
       adsurl = {https://ui.adsabs.harvard.edu/abs/2023MNRAS.524.4907S}
}

@ARTICLE{Giavalisco11,
       author = {{Giavalisco}, Mauro and {Vanzella}, Eros and {Salimbeni}, Sara and {Tripp}, Todd M. and {Dickinson}, Mark and {Cassata}, Paolo and {Renzini}, Alvio and {Guo}, Yicheng and {Ferguson}, Henry C. and {Nonino}, Mario and {Cimatti}, Andrea and {Kurk}, Jaron and {Mignoli}, Marco and {Tang}, Yuping},
        title = "{Discovery of Cold, Pristine Gas Possibly Accreting onto an Overdensity of Star-forming Galaxies at Redshift z \raisebox{-0.5ex}\textasciitilde 1.6}",
      journal = {\apj},
         year = 2011,
        month = dec,
       volume = {743},
       number = {1},
          eid = {95},
        pages = {95},
          doi = {10.1088/0004-637X/743/1/95},
archivePrefix = {arXiv},
       eprint = {1106.1205},
 primaryClass = {astro-ph.CO},
       adsurl = {https://ui.adsabs.harvard.edu/abs/2011ApJ...743...95G}
}

@ARTICLE{Ribaudo11,
       author = {{Ribaudo}, Joseph and {Lehner}, Nicolas and {Howk}, J. Christopher and {Werk}, Jessica K. and {Tripp}, Todd M. and {Prochaska}, J. Xavier and {Meiring}, Joseph D. and {Tumlinson}, Jason},
        title = "{Evidence for Cold Accretion: Primitive Gas Flowing onto a Galaxy at z \raisebox{-0.5ex}\textasciitilde 0.274}",
      journal = {\apj},
         year = 2011,
        month = dec,
       volume = {743},
       number = {2},
          eid = {207},
        pages = {207},
          doi = {10.1088/0004-637X/743/2/207},
archivePrefix = {arXiv},
       eprint = {1105.5381},
 primaryClass = {astro-ph.CO},
       adsurl = {https://ui.adsabs.harvard.edu/abs/2011ApJ...743..207R}
}

@ARTICLE{Rubin12,
       author = {{Rubin}, Kate H.~R. and {Prochaska}, J. Xavier and {Koo}, David C. and {Phillips}, Andrew C.},
        title = "{The Direct Detection of Cool, Metal-enriched Gas Accretion onto Galaxies at z \raisebox{-0.5ex}\textasciitilde 0.5}",
      journal = {\apjl},
         year = 2012,
        month = mar,
       volume = {747},
       number = {2},
          eid = {L26},
        pages = {L26},
          doi = {10.1088/2041-8205/747/2/L26},
archivePrefix = {arXiv},
       eprint = {1110.0837},
 primaryClass = {astro-ph.CO},
       adsurl = {https://ui.adsabs.harvard.edu/abs/2012ApJ...747L..26R}
}

@ARTICLE{Shen14,
       author = {{Shen}, Sijing and {Madau}, Piero and {Conroy}, Charlie and {Governato}, Fabio and {Mayer}, Lucio},
        title = "{The Baryon Cycle of Dwarf Galaxies: Dark, Bursty, Gas-rich Polluters}",
      journal = {\apj},
         year = 2014,
        month = sep,
       volume = {792},
       number = {2},
          eid = {99},
        pages = {99},
          doi = {10.1088/0004-637X/792/2/99},
archivePrefix = {arXiv},
       eprint = {1308.4131},
 primaryClass = {astro-ph.CO},
       adsurl = {https://ui.adsabs.harvard.edu/abs/2014ApJ...792...99S}
}

@ARTICLE{Christensen18,
       author = {{Christensen}, Charlotte R. and {Dav{\'e}}, Romeel and {Brooks}, Alyson and {Quinn}, Thomas and {Shen}, Sijing},
        title = "{Tracing Outflowing Metals in Simulations of Dwarf and Spiral Galaxies}",
      journal = {\apj},
         year = 2018,
        month = nov,
       volume = {867},
       number = {2},
          eid = {142},
        pages = {142},
          doi = {10.3847/1538-4357/aae374},
archivePrefix = {arXiv},
       eprint = {1808.07872},
 primaryClass = {astro-ph.GA},
       adsurl = {https://ui.adsabs.harvard.edu/abs/2018ApJ...867..142C}
}

@ARTICLE{McQuinn15,
       author = {{McQuinn}, Kristen B.~W. and {Skillman}, Evan D. and {Dolphin}, Andrew and {Cannon}, John M. and {Salzer}, John J. and {Rhode}, Katherine L. and {Adams}, Elizabeth A.~K. and {Berg}, Danielle and {Giovanelli}, Riccardo and {Haynes}, Martha P.},
        title = "{Leo P: How Many Metals Can a Very Low Mass, Isolated Galaxy Retain?}",
      journal = {\apjl},
         year = 2015,
        month = dec,
       volume = {815},
       number = {2},
          eid = {L17},
        pages = {L17},
          doi = {10.1088/2041-8205/815/2/L17},
archivePrefix = {arXiv},
       eprint = {1512.00459},
 primaryClass = {astro-ph.GA},
       adsurl = {https://ui.adsabs.harvard.edu/abs/2015ApJ...815L..17M}
}

@ARTICLE{Qu22,
       author = {{Qu}, Zhijie and {Bregman}, Joel N.},
        title = "{Absorption Line Search through Three Local Group Dwarf Galaxy Halos}",
      journal = {\apj},
         year = 2022,
        month = mar,
       volume = {927},
       number = {2},
          eid = {228},
        pages = {228},
          doi = {10.3847/1538-4357/ac51df},
archivePrefix = {arXiv},
       eprint = {2203.08246},
 primaryClass = {astro-ph.GA},
       adsurl = {https://ui.adsabs.harvard.edu/abs/2022ApJ...927..228Q}
}

@ARTICLE{Zheng24,
       author = {{Zheng}, Yong and {Faerman}, Yakov and {Oppenheimer}, Benjamin D. and {Putman}, Mary E. and {McQuinn}, Kristen B.~W. and {Kirby}, Evan N. and {Burchett}, Joseph N. and {Telford}, O. Grace and {Werk}, Jessica K. and {Kim}, Doyeon A.},
        title = "{A Comprehensive Investigation of Metals in the Circumgalactic Medium of Nearby Dwarf Galaxies}",
      journal = {\apj},
         year = 2024,
        month = jan,
       volume = {960},
       number = {1},
          eid = {55},
        pages = {55},
          doi = {10.3847/1538-4357/acfe6b},
archivePrefix = {arXiv},
       eprint = {2301.12233},
 primaryClass = {astro-ph.GA},
       adsurl = {https://ui.adsabs.harvard.edu/abs/2024ApJ...960...55Z}
}

@ARTICLE{Lelli12,
       author = {{Lelli}, F. and {Verheijen}, M. and {Fraternali}, F. and {Sancisi}, R.},
        title = "{Dynamics of starbursting dwarf galaxies: I Zw 18}",
      journal = {\aap},
         year = 2012,
        month = jan,
       volume = {537},
          eid = {A72},
        pages = {A72},
          doi = {10.1051/0004-6361/201117867},
archivePrefix = {arXiv},
       eprint = {1110.6042},
 primaryClass = {astro-ph.CO},
       adsurl = {https://ui.adsabs.harvard.edu/abs/2012A&A...537A..72L}
}

@ARTICLE{Lopez12,
       author = {{L{\'o}pez-S{\'a}nchez}, {\'A}. R. and {Koribalski}, B.~S. and {van Eymeren}, J. and {Esteban}, C. and {Kirby}, E. and {Jerjen}, H. and {Lonsdale}, N.},
        title = "{The intriguing H I gas in NGC 5253: an infall of a diffuse, low-metallicity H I cloud?}",
      journal = {\mnras},
         year = 2012,
        month = jan,
       volume = {419},
       number = {2},
        pages = {1051-1069},
          doi = {10.1111/j.1365-2966.2011.19762.x},
archivePrefix = {arXiv},
       eprint = {1109.0806},
 primaryClass = {astro-ph.CO},
       adsurl = {https://ui.adsabs.harvard.edu/abs/2012MNRAS.419.1051L}
}

@ARTICLE{Stil02,
       author = {{Stil}, J.~M. and {Israel}, F.~P.},
        title = "{HI distribution and kinematics of NGC 1569}",
      journal = {\aap},
         year = 2002,
        month = sep,
       volume = {392},
        pages = {473-487},
          doi = {10.1051/0004-6361:20020953},
archivePrefix = {arXiv},
       eprint = {astro-ph/0209444},
 primaryClass = {astro-ph},
       adsurl = {https://ui.adsabs.harvard.edu/abs/2002A&A...392..473S}
}

@ARTICLE{Chandola23,
       author = {{Chandola}, Y. and {Tsai}, C.~W. and {Li}, D. and {Sengupta}, C. and {Ma}, Y.~Z. and {Zuo}, P.},
        title = "{GMRT H I mapping of mid-infrared bright blue compact dwarf galaxies W1016+3754 and W2326+0608}",
      journal = {\mnras},
         year = 2023,
        month = aug,
       volume = {523},
       number = {3},
        pages = {3848-3862},
          doi = {10.1093/mnras/stad1618},
archivePrefix = {arXiv},
       eprint = {2305.16786},
 primaryClass = {astro-ph.GA},
       adsurl = {https://ui.adsabs.harvard.edu/abs/2023MNRAS.523.3848C}
}

@ARTICLE{Kobulnicky08,
       author = {{Kobulnicky}, Henry A. and {Skillman}, Evan D.},
        title = "{Inflows and Outflows in the Dwarf Starburst Galaxy NGC 5253: High-Resolution H I Observations}",
      journal = {\aj},
         year = 2008,
        month = feb,
       volume = {135},
       number = {2},
        pages = {527-537},
          doi = {10.1088/0004-6256/135/2/527},
archivePrefix = {arXiv},
       eprint = {0711.2688},
 primaryClass = {astro-ph},
       adsurl = {https://ui.adsabs.harvard.edu/abs/2008AJ....135..527K}
}

@ARTICLE{Cannon04,
       author = {{Cannon}, John M. and {McClure-Griffiths}, N.~M. and {Skillman}, Evan D. and {C{\^o}t{\'e}}, St{\'e}phanie},
        title = "{The Complex Neutral Gas Dynamics of the Dwarf Starburst Galaxy NGC 625}",
      journal = {\apj},
         year = 2004,
        month = may,
       volume = {607},
       number = {1},
        pages = {274-284},
          doi = {10.1086/383408},
archivePrefix = {arXiv},
       eprint = {astro-ph/0403190},
 primaryClass = {astro-ph},
       adsurl = {https://ui.adsabs.harvard.edu/abs/2004ApJ...607..274C}
}

@ARTICLE{Lelli14b,
       author = {{Lelli}, Federico and {Verheijen}, Marc and {Fraternali}, Filippo},
        title = "{Dynamics of starbursting dwarf galaxies. III. A H I study of 18 nearby objects}",
      journal = {\aap},
         year = 2014,
        month = jun,
       volume = {566},
          eid = {A71},
        pages = {A71},
          doi = {10.1051/0004-6361/201322657},
archivePrefix = {arXiv},
       eprint = {1404.6252},
 primaryClass = {astro-ph.GA},
       adsurl = {https://ui.adsabs.harvard.edu/abs/2014A&A...566A..71L}
}

@ARTICLE{Kimbro21,
       author = {{Kimbro}, Erin and {Reines}, Amy E. and {Molina}, Mallory and {Deller}, Adam T. and {Stern}, Daniel},
        title = "{Clumpy Star Formation and AGN Activity in the Dwarf-Dwarf Galaxy Merger Mrk 709}",
      journal = {\apj},
         year = 2021,
        month = may,
       volume = {912},
       number = {2},
          eid = {89},
        pages = {89},
          doi = {10.3847/1538-4357/abec6a},
archivePrefix = {arXiv},
       eprint = {2103.05655},
 primaryClass = {astro-ph.GA},
       adsurl = {https://ui.adsabs.harvard.edu/abs/2021ApJ...912...89K}
}

@ARTICLE{Egorov21,
       author = {{Egorov}, Oleg V. and {Lozinskaya}, Tatiana A. and {Vasiliev}, Konstantin I. and {Yarovova}, Anastasiya D. and {Gerasimov}, Ivan S. and {Kreckel}, Kathryn and {Moiseev}, Alexei V.},
        title = "{Star formation in the nearby dwarf galaxy DDO 53: interplay between gas accretion and stellar feedback}",
      journal = {\mnras},
         year = 2021,
        month = dec,
       volume = {508},
       number = {2},
        pages = {2650-2667},
          doi = {10.1093/mnras/stab2710},
archivePrefix = {arXiv},
       eprint = {2109.08526},
 primaryClass = {astro-ph.GA},
       adsurl = {https://ui.adsabs.harvard.edu/abs/2021MNRAS.508.2650E}
}

@ARTICLE{Tang22,
       author = {{Tang}, Yimeng and {Tao}, Bojun and {Zhang}, Hong-Xin and {Chen}, Guangwen and {Gao}, Yulong and {Lin}, Zesen and {Yao}, Yao and {Shi}, Yong and {Kong}, Xu},
        title = "{Unveiling the formation of NGC 2915 with MUSE: A counter-rotating stellar disk embedded in a disordered gaseous environment}",
      journal = {\aap},
         year = 2022,
        month = dec,
       volume = {668},
          eid = {A179},
        pages = {A179},
          doi = {10.1051/0004-6361/202243944},
archivePrefix = {arXiv},
       eprint = {2210.06485},
 primaryClass = {astro-ph.GA},
       adsurl = {https://ui.adsabs.harvard.edu/abs/2022A&A...668A.179T}
}

@ARTICLE{Chhatkuli23,
       author = {{Chhatkuli}, Daya Nidhi and {Paudel}, Sanjaya and {Bachchan}, Rajesh Kumar and {Aryal}, Binil and {Yoo}, Jaewon},
        title = "{Forming blue compact dwarf galaxy through mergers}",
      journal = {\mnras},
         year = 2023,
        month = apr,
       volume = {520},
       number = {4},
        pages = {4953-4960},
          doi = {10.1093/mnras/stac3700},
archivePrefix = {arXiv},
       eprint = {2212.07022},
 primaryClass = {astro-ph.GA},
       adsurl = {https://ui.adsabs.harvard.edu/abs/2023MNRAS.520.4953C}
}

@ARTICLE{Silk87,
       author = {{Silk}, Joseph and {Wyse}, Rosemary F.~G. and {Shields}, G.~A.},
        title = "{On the Origin of Dwarf Galaxies}",
      journal = {\apjl},
         year = 1987,
        month = nov,
       volume = {322},
        pages = {L59},
          doi = {10.1086/185037},
       adsurl = {https://ui.adsabs.harvard.edu/abs/1987ApJ...322L..59S}
}

@ARTICLE{Elmegreen12,
       author = {{Elmegreen}, Bruce G. and {Zhang}, Hong-Xin and {Hunter}, Deidre A.},
        title = "{In-spiraling Clumps in Blue Compact Dwarf Galaxies}",
      journal = {\apj},
         year = 2012,
        month = mar,
       volume = {747},
       number = {2},
          eid = {105},
        pages = {105},
          doi = {10.1088/0004-637X/747/2/105},
archivePrefix = {arXiv},
       eprint = {1201.3658},
 primaryClass = {astro-ph.CO},
       adsurl = {https://ui.adsabs.harvard.edu/abs/2012ApJ...747..105E}
}

@ARTICLE{Bekki08,
       author = {{Bekki}, Kenji},
        title = "{Formation of blue compact dwarf galaxies from merging and interacting gas-rich dwarfs}",
      journal = {\mnras},
         year = 2008,
        month = jul,
       volume = {388},
       number = {1},
        pages = {L10-L14},
          doi = {10.1111/j.1745-3933.2008.00489.x},
archivePrefix = {arXiv},
       eprint = {0805.0662},
 primaryClass = {astro-ph},
       adsurl = {https://ui.adsabs.harvard.edu/abs/2008MNRAS.388L..10B}
}

@ARTICLE{Hunter04,
       author = {{Hunter}, Deidre A. and {Elmegreen}, Bruce G.},
        title = "{Star Formation Properties of a Large Sample of Irregular Galaxies}",
      journal = {\aj},
         year = 2004,
        month = nov,
       volume = {128},
       number = {5},
        pages = {2170-2205},
          doi = {10.1086/424615},
archivePrefix = {arXiv},
       eprint = {astro-ph/0408229},
 primaryClass = {astro-ph},
       adsurl = {https://ui.adsabs.harvard.edu/abs/2004AJ....128.2170H}
}

@ARTICLE{Putman21,
       author = {{Putman}, Mary E. and {Zheng}, Yong and {Price-Whelan}, Adrian M. and {Grcevich}, Jana and {Johnson}, Amalya C. and {Tollerud}, Erik and {Peek}, Joshua E.~G.},
        title = "{The Gas Content and Stripping of Local Group Dwarf Galaxies}",
      journal = {\apj},
         year = 2021,
        month = may,
       volume = {913},
       number = {1},
          eid = {53},
        pages = {53},
          doi = {10.3847/1538-4357/abe391},
archivePrefix = {arXiv},
       eprint = {2101.07809},
 primaryClass = {astro-ph.GA},
       adsurl = {https://ui.adsabs.harvard.edu/abs/2021ApJ...913...53P}
}

@ARTICLE{DiTeodoro15,
       author = {{Di Teodoro}, E.~M. and {Fraternali}, F.},
        title = "{$^{3D}$ BAROLO: a new 3D algorithm to derive rotation curves of galaxies}",
      journal = {\mnras},
         year = 2015,
        month = aug,
       volume = {451},
       number = {3},
        pages = {3021-3033},
          doi = {10.1093/mnras/stv1213},
archivePrefix = {arXiv},
       eprint = {1505.07834},
 primaryClass = {astro-ph.GA},
       adsurl = {https://ui.adsabs.harvard.edu/abs/2015MNRAS.451.3021D}
}
\bibliographystyle{aasjournal}

\end{CJK*}
\end{document}